\documentclass[journal]{IEEEtran}

\ifCLASSINFOpdf
\else
\fi

\usepackage{setspace}
\usepackage{duckuments} % Load the package for the yellow duck placeholder
\usepackage{times}
\usepackage{subfig}
\usepackage{graphicx}
\usepackage{amsmath,amsthm}
\usepackage{amssymb,wasysym}
\usepackage{mathrsfs}
\usepackage{cite}
\usepackage{xcolor}
\usepackage{url}
\usepackage[lined,linesnumbered,ruled]{algorithm2e}
\usepackage{wrapfig}
\usepackage{verbatim}
\usepackage{dsfont}
\usepackage{algorithmic}
\usepackage{multirow}

\ifodd 0
\newcommand{\rev}[1]{{\color{green}#1}} %revise of the text
\newcommand{\newrev}[1]{{\color{blue}#1}} %revise of the text
\newcommand{\needrev}[1]{{\color{red}#1}} %revise of the text
\newcommand{\nextrev}[1]{{\color{orange}#1}} 
\newcommand{\waitrev}[1]{{\color{pink}#1}} 
\else
\newcommand{\rev}[1]{#1}
\newcommand{\newrev}[1]{#1} %revise of the text
\newcommand{\needrev}[1]{#1} %revise of the text
\newcommand{\nextrev}[1]{#1} %revise of the text
\newcommand{\waitrev}[1]{#1} %revise of the text
\fi

\begin{document}

\newcommand{\name}{DUAL-Health\xspace}
% FIND-Health: Fusion-based IN-car Dynamic Health monitoring
% AURA: Adaptive Uncertainty-aware outdooR heAlth monitoring
% FIRM-Health (Flexible Integration for Robust Multimodal health monitoring)
% FUSE-Health: Flexible Uncertainty-aware Sensing/System in dynamic Environments for Health monitoring
% Dynamic Uncertainty-Aware Learning for multimodal health monitoring
% HEART (Health monitoring with uncErtainty quAntification and multimodal fuRsion for ouTdoor environments)

\title{Dynamic Uncertainty-aware Multimodal Fusion for Outdoor Health Monitoring}

\author{Zihan Fang, Zheng Lin, Senkang Hu, Yihang Tao, Yiqin Deng,~\IEEEmembership{Member,~IEEE}, \\
Xianhao Chen,~\IEEEmembership{Member,~IEEE} and Yuguang Fang,~\IEEEmembership{Fellow,~IEEE}

\thanks{{The research work described in this paper was conducted in the JC STEM Lab of Smart City funded by The Hong Kong Jockey Club Charities Trust under Contract 2023-0108.  This work was also supported in part by the Hong Kong SAR Government under the Global STEM Professorship and Research Talent Hub, and in part by the Hong Kong Innovation and Technology Commission under InnoHK Project CIMDA. The work of Yiqin Deng was supported in part by the National Natural Science Foundation of China under Grant No. 62301300. The work of Xianhao Chen was supported in part by HKU-SCF FinTech Academy R\&D Funding.}
% \textit{(Corresponding author: Yiqin Deng)}
}

\thanks{Z. Fang, S. Hu, Y. Tao, Y. Deng and Y. Fang are with Hong Kong JC STEM Lab of Smart City and Department of Computer Science, City University of Hong Kong, Kowloon, Hong Kong SAR, China (e-mail: zihanfang3-c@my.cityu.edu.hk; senkang.forest@my.cityu.edu.hk; yihang.tommy@my.cityu.edu.hk; yiqideng@cityu.edu.hk; my.fang@cityu.edu.hk).}
\thanks{Z. Lin and X. Chen are with the Department of Electrical and Electronic Engineering, The University of Hong Kong, Pok Fu Lam, Hong Kong, China (e-mail: linzheng@eee.hku.hk; xchen@eee.hku.hk).}

}

%\markboth{Journal of \LaTeX\ Class Files,~Vol.~14, No.~8, August~2015}% {Shell \MakeLowercase{\textit{et al.}}: Bare Advanced Demo of IEEEtran.cls for IEEE Computer Society Journals}

\maketitle

\begin{abstract}
% 流行的户外健康监方法通过多模态传感器指标来检测易受影响的人群（如老年人）的健康状态从而实现健康问题的实时检测。
Outdoor health monitoring is essential to detect early abnormal health status for safeguarding human health and safety. 
% for safeguarding vulnerable populations, such as drivers and the elderly.
% By enabling early detection of abnormal health biomarkers (e.g., irregular heart rates or behavioral changes) through multimodal data fusion, outdoor health monitoring facilitates timely interventions to prevent health deterioration.
Conventional outdoor monitoring relies on static multimodal deep learning frameworks, which requires extensive data training from scratch and fails to capture subtle health status changes. Multimodal large language models (MLLMs) emerge as a promising alternative, utilizing only small datasets to fine-tune pre-trained information-rich models for enabling powerful health status monitoring.
% While traditional multimodal fusion methods are limited by data scarcity due to the rarity of abnormal data \rev{and annotation expertise}, 
% multimodal large language models (MLLMs) emerge as a promising solution by leveraging pre-trained medical knowledge for fine-tuning with small datasets and large number of parameters to recognize subtle health changes.
% Their large number of parameters also enables the learning of complex health patterns for the timely recognition of subtle health changes.
% with the benefits of pre-trained medical knowledge to fine-tune model with small datasets and the substantial number of parameters to learn complex health patterns, allowing for timely recognition of subtle health changes.
% (e.g., fluctuating weather and changing traffic conditions)
\waitrev{Unfortunately, MLLM-based outdoor health monitoring also faces significant challenges: i) sensor data contains input noise stemming from sensor data acquisition 
% (e.g., external environmental changes and mobility of the individuals), 
and fluctuation noise caused by sudden changes in physiological signals due to dynamic outdoor environments, 
% such as \rev{lighting and temperature changes}, 
thus degrading the training performance; ii) current transformer-based MLLMs struggle to achieve robust multimodal fusion, as they lack a design for fusing the noisy modality}; \rev{iii) modalities with varying noise levels hinder accurate recovery of missing data from fluctuating distributions.}
% This limitation disrupts feature alignment and hampers the detection of subtle health changes. 
% Moreover, current transformer-based MLLMs struggle with capturing critical cross-modal relationships, as they often fail to account for the varying quality of modalities, which disrupts feature alignment and hampers the detection of subtle health changes.
% it is hard to separate the data uncertainty caused by quality degradation from the high-variance fluctuations in health biomarkers, often leading to delays in recognizing critical events. 
% 此外，噪声的低质量模态破坏了多模态融合进行健康状态评估的特征对齐，从而降低跨模态融合后的健康变化的检测精度。
% undermining the capture of cross-modal correlation
To combat these challenges, we propose an uncertainty-aware multimodal fusion framework, named \name, 
% which consists two key components to accurately quantify uncertainty and strength cross-modal fusion 
for outdoor health monitoring in dynamic and noisy environments. First, to assess the impact of noise, we accurately quantify modality uncertainty caused by input and fluctuation noise with current and temporal features. 
% separate noisy inputs from meaningful health \rev{biomarker} fluctuations, we accurately quantify uncertainty caused by data degradation and \rev{biomarker} variability with current and temporal features. 
Second, to empower efficient muitimodal fusion with low-quality modalities, we customize the fusion weight for each modality based on quantified and calibrated uncertainty.
\rev{Third, to enhance data recovery from fluctuating noisy modalities, we align modality distributions within a common semantic space.}
% transfer the distributions of different modalities into a common semantic space
% , showcasing its robustness in dynamical environments with vary data quality.
% , addressing the challenges of uncertainty and cross-modal misalignment in real-world health monitoring scenarios.
Extensive experiments demonstrate that our \name outperforms state-of-the-art baselines in detection accuracy and robustness.
\end{abstract}


\IEEEpeerreviewmaketitle

\vspace{-0.2cm}
\section{Introduction} \label{sec:introduction}
\begin{figure}[t]
\centering
\includegraphics[width=0.9\linewidth]{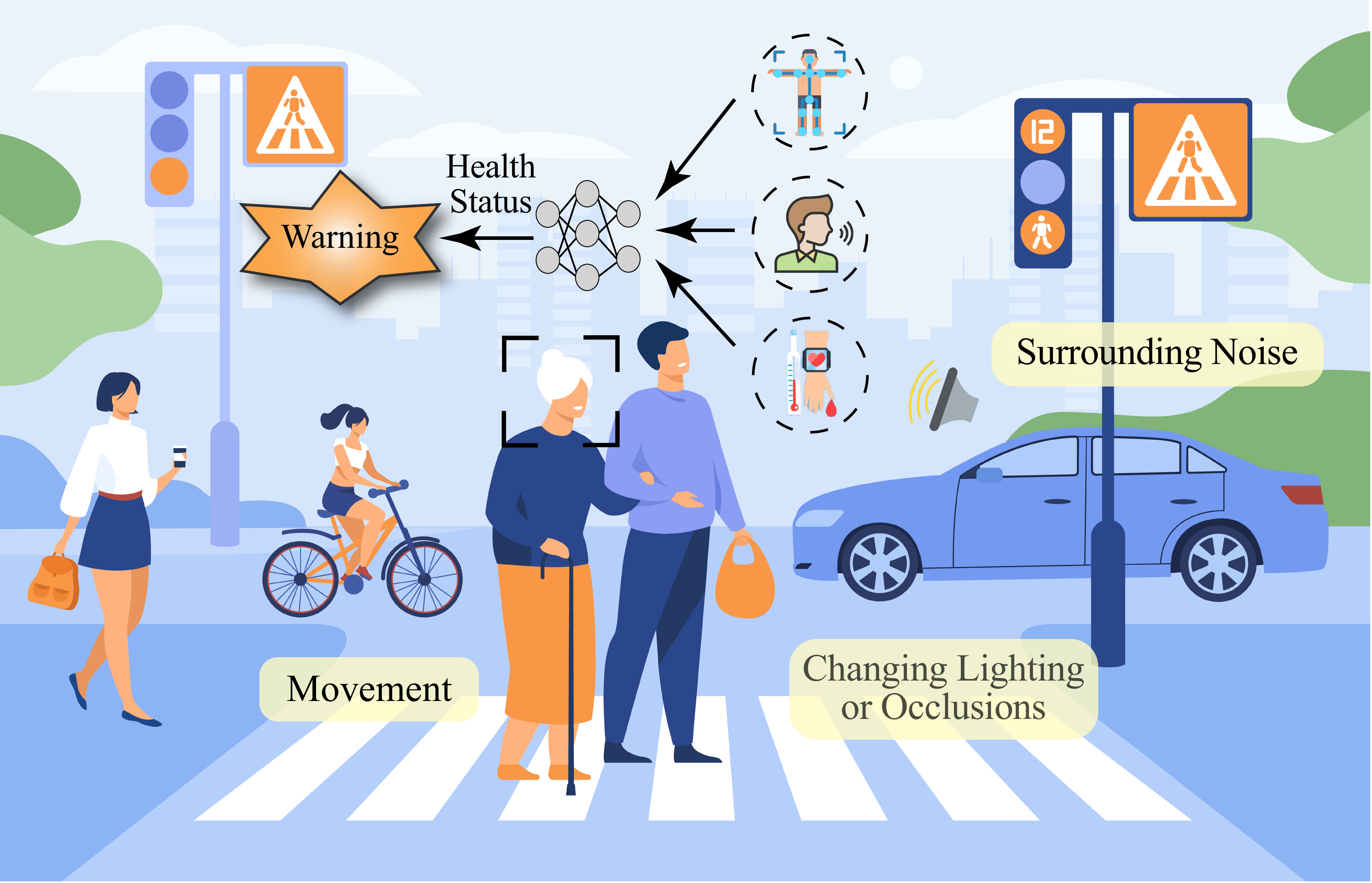}
\caption{The typical scenario of outdoor health monitoring with the integration of multimodal data.}
\label{fig:scenario_challenge}
\vspace{-2ex}
\end{figure}

Cardiovascular diseases are the leading cause of death globally, accounting for approximately 17.9 million deaths annually~\cite{WHO, wang2016global}. According to the World Health Organization~\cite{WHO}, 85\% of these deaths are attributed to heart attacks and strokes, many of which occur outdoors or in non-clinical settings.
Outdoor health monitoring plays a crucial role in safeguarding people's health and public safety, such as enabling real-time detection of potential health issues like cardiovascular disease and stroke among drivers and the elderly~\cite{hamza2020monitoring, qin2020imaging,fang2024ic3m}.
By early detection of abnormal health biomarkers, such as irregular heart rates or behavioral changes, outdoor health monitoring allows for timely interventions to prevent health crises and ensure prompt medical attention~\cite{al2019remote,tang2024merit}, \nextrev{making it a vital research area for public safety}.
% Outdoor health monitoring holds great potential to identify early signs of health deterioration,  timely intervention to prevent health crises and provide prompt medical assistance for elderly individuals, therefore, making it a vital research area to ensure elderly safety~\cite{}.
% 
% 
% 
% To enhance the reliability of the early detection for potential health issue, diverse data from various sensors (e.g., xxx) are integrated to provide complementary information from different perspectives.
% Outdoor health monitoring \rev{holds great potential to identify early signs of health deterioration} and has emerged as a vital research area.
The complex and dynamic nature of outdoor environments necessitates the integration of multimodal data from diverse sensors, including physiological signals~\cite{li2024llava, moor2023med, xu2018raim}, facial expressions~\cite{yang2022disentangled, lv2021progressive}, speech patterns~\cite{akbari2021vatt, shi2022learning, mei2024wavcaps}, and self-reported measurements~\cite{gao2020compose, zhang2022m3care}.
% By leveraging multimodal fusion techniques to capture physiological and behavioral health biomarkers in different perspectives, abnormal health event can be detected and trigger automated intervention to \waitrev{release} alerts or contact emergency services, thereby ensuring timely medical assistance and reducing the risk of health deterioration. 
As depicted in Fig.~\ref{fig:scenario_challenge}, \nextrev{multimodal fusion methods harness complementary information from diverse data modalities, enabling the reliable detection of abnormal health biomarkers \rev{for automated interventions to alert or contact emergency services~\cite{hu2025cp}.}}

% 在健康医疗领域，高质量标签数据很少->从头训练多模态模型难，原因，影响
Unfortunately, traditional multimodal fusion methods~\cite{yang2022disentangled, lv2021progressive, akbari2021vatt, shi2022learning, mei2024wavcaps, gao2020compose, zhang2022m3care} typically require massive data to learn the complex and diverse patterns underlying health biomarkers from scratch, while the scarcity of task-specific data severely limits their generalization ability. Due to the rarity and unpredictability of abnormal health status~\cite{bao2024bmad} and the specialized expertise required for accurate annotation~\cite{verspoor2012corpus},
\nextrev{this data scarcity restricts the model’s ability to learn representative patterns, leading to poor generalization to unseen or out-of-distribution data~\cite{chan2024medtsllm}.}
% \waitrev{The scarcity of labeled data limits models' ability to uncover underlying feature relationships and capture discriminative features for subtle health changes.}
% 不能覆盖全部的样本空间，在未见过的数据上表现差
% 预训练大模型可以few-shot有效学习到数据的潜在模式，多模态大模型有潜力 (motivation) 
Recent advancements in multimodal large language models (MLLMs) have already shown their potential for health monitoring~\cite{chan2024medtsllm,lin2023pushing,li2024llava,hu2025collaborative,lin2025hsplitlora,moor2023med, kim2024health,fang2024automated,lin2024split,liu2023large,hu2025task}. 
% Pre-trained on extensive datasets, MLLMs possess general medical knowledge and strong transfer learning capabilities, enabling quick adaptation to specific healthcare tasks with minimal labeled data during fine-tuning~\cite{thirunavukarasu2023large, yin2023survey, wang2024augmented}.  
Unlike traditional methods, MLLMs are pre-trained on extensive and diverse datasets, allowing them to acquire broad medical knowledge.
Instead of learning from scratch, these models leverage their pre-trained knowledge and require only minimal labeled data for fine-tuning specific healthcare tasks~\cite{thirunavukarasu2023large, yin2023survey, wang2024augmented,lin2024splitlora}, significantly reducing the dependence on large task-specific datasets.
% Moreover, the substantial number of parameters in MLLMs further increases their ability to identify subtle changes in health biomarkers~\cite{chan2024medtsllm}, facilitating timely interventions through early detection of health issues.
% traditional fusion methods rely heavily on handcrafted feature extracting and separate feature extracting
% Moreover, MLLMs inherently learn cross-modal representations during pre-training, enabling it to capture subtle correlations and patterns across modalities, which are often overlooked by traditional methods.
Moreover, the substantial number of parameters allows MLLMs to model highly detailed representations of multimodal inputs, further enhancing sensitivity to subtle changes in health biomarkers~\cite{chan2024medtsllm}, such as slight changes in heart rate or facial expressions. 
\rev{This capability facilitates early detection of abnormal health status and timely interventions.}
% (e.g., gradual changes in vital signs that may indicate a risk of heart failure or stroke).
% With their capacity to embed different modalities into a common feature space and focus on relevant features through attention mechanisms, multimodal LLMs (MLLMs) in health event detection~\cite{} have gained immense popularity recently. 

% 问题 - (车内监测更容易受到干扰，导致biomarker波动，误判) 车辆行驶的动态环境变化会影响多模态数据的质量，多模态模型效果变差
Though many MLLM frameworks~\cite{chan2024medtsllm, hu2025agentscomerge,li2024llava, moor2023med, kim2024health, liu2023large,hu2024agentscodriver} have made significant strides in health monitoring, implementing MLLMs for outdoor health monitoring is non-trivial.
% the \newrev{changing external environments often degrade data quality across modalities~\cite{wang2021m, nicolaou2011continuous}}, bringing new challenges for timely health status detection. 
First, accurately quantifying modality uncertainty under dynamic noise environments is highly challenging.
Outdoor health monitoring performance is often degraded due to data uncertainty stemming from \rev{environmental} noise. On the one hand, sensors may introduce input noise from data acquisition due to environmental changes \rev{(e.g., sudden shadows and mobility of individuals), where multiple types of noise can simultaneously affect the same modality.} 
% \waitrev{On the other hand, dynamic environmental changes can trigger sudden \rev{non-pathological} fluctuations in physiological signals \needrev{which may indicate potential health emergencies in the monitored individuals}, referred to as fluctuation noise. For instance, sudden traffic accidents may induce stress or emotional shifts, leading to abrupt physiological signal changes such as spikes in heart rate or variations in respiration. 
% \needrev{However, the intertwined effects of input and fluctuation noises complicates the quantification of modality uncertainty for multimodal fusion, making these fluctuations easily misclassified as noise rather than early biomarkers of health issues, thus resulting in false alarms or missing detections of abnormal health status.}}
% % These fluctuations closely resemble early biomarkers of health issues, making it challenging to distinguish between noise-induced physiological signal changes and actual health problems.
\waitrev{On the other hand, dynamic environmental changes may trigger health issues, such as cardiovascular emergencies, which often present early biomarkers before health deterioration, including fluctuations in physiological signals. For instance, sudden traffic accidents may induce stress or emotional shifts, leading to abrupt physiological changes such as spikes in heart rate or variations in respiration. These fluctuations closely resemble input noise, referred to as fluctuation noise, making them easily misclassified as noise rather than early biomarkers of health issues. This intertwined effects of input and fluctuation noises complicates the quantification of modality uncertainty, resulting in false alarms or missing detections of abnormal health status.}

Second, achieving robust multimodal fusion under complicated noise conditions remains challenging, as it requires \needrev{accurate estimation of each modality’s reliability} and dynamic adjustment of their contributions.
\waitrev{Noise in dynamic environments varies over time and diverse modalities exhibit different sensitivities. For example, visual data is susceptible to disruptions from lighting variations, shadows, or occlusions, while audio data may be masked by background noise such as traffic or wind.}
% Current MLLMs~\cite{kim2024health, moor2023med, chan2024medtsllm} often extract cross-modal feature relationships through the self-attention mechanism, disregarding the impact of low-quality modalities on critical relationship extraction during cross-modal interactions.
However, current MLLMs~\cite{chan2024medtsllm, li2024llava, moor2023med, kim2024health, liu2023large} often treat all modalities as if they contribute equally to a task, failing to account for their varying data quality.
This uniform treatment diminishes the model’s ability to distinguish noisy inputs from meaningful features~\cite{wu2024noiseboost},
% suppress noisy inputs and prioritize informative features~\cite{wu2024noiseboost}. 
% However, current MLLMs~\cite{chan2024medtsllm, li2024llava, moor2023med, kim2024health, liu2023large} often fail to account for the detrimental impact of low-quality noisy modalities during the multimodal fusion. These models treat all modalities as if they contribute equally to a task, regardless of their qualities.
% This disruption misleads the attention mechanism to focus on insignificant or irrelevant feature relationships, thereby posing significant barriers to identify subtle changes in health biomarkers and thus leading to missed detection.
misleading the model's attention to focus on irrelevant features, which can obscure critical modalities and lead to missed detections or incorrect predictions. 
% \rev{As a result, all these issues underscore the necessity} to precisely estimate the reliability and sensitivity of each multimodal sample to achieve adaptive transformer-based feature fusion in dynamic changing driving environments.
Therefore, designing dynamic weighting strategy to adjust contributions for multimodal fusion, especially within transformer-based architectures, remains a critical issue.

Finally, recovering missing data from available modalities of varying quality is challenging.
\waitrev{Dynamic environments also cause modality missing, such as pedestrian occlusions disrupting camera inputs or body posture changes affecting physiological signals, necessitating reliable information from remaining modalities to recover missing data and mitigate performance degradation~\cite{wang2020transmodality, wang2020multimodal, zhao2021missing, zhou2021memorizing, perera2019ocgan}.}
% Existing data imputation approaches always assume that the remaining modalities are of high quality and leverage their stable patterns to recover the missing data.}
% % leveraging the auxiliary data to improve the reliability of results.
% % sensor readings may exhibit drift or noise under different environmental conditions. This leads to temporal or contextual shifts in modality-specific data distributions.
% % as dynamic environment exacerbates the inconsistency in distributions across modalities.
% However, dynamic changes in data quality \rev{exacerbate} discrepancies between modalities, leading to fluctuating modality distributions that disrupt the stable cross-modal alignment.
However, in dynamic environments, the data quality of available modalities changes over time due to varying noise levels, making the data distributions of modalities fluctuate dynamically.
This instability hinders consistent cross-modal alignment, making it difficult to capture stable semantic correlations and reliable complementary information, resulting in inaccurate data recovery that fails to capture critical details and compromises detection accuracy.
% These distribution fluctuations disrupt stable cross-modal alignment, hindering the capture of consistent semantic correlations. As a result, models struggle to effectively leverage complementary information from available modalities, resulting in inaccurate data recovery that fails to capture critical details and compromises detection accuracy.

% 核心思想和贡献
% 我们提出一种动态调整权重计算策略，基于动态驾驶环境中变化的数据质量和模态丢失，从而提升健康状态监测的实时性和有效性。
In this paper, we propose a dynamic multimodal fusion framework, \needrev{named \underline{D}ynamic \underline{U}ncertainty-\underline{A}ware \underline{L}earning (\name)}, for outdoor health monitoring in dynamic and noisy environments.
% 为了准确表征数据的不确定性，我们估计单个输入数据的不确定性，降低噪声对特征提取的影响，同时，捕捉数据随时间变化的波动性和趋势，在状态波动剧烈时，增加对高方差模态的关注。
% First, to accurately represent data uncertainty for adaptive assignment of modality-specific weights, 
The \name framework consists of three key components: modality uncertainty quantification, transformer-based multimodal fusion, and missing modality reconstruction. The modality uncertainty quantification utilizes current and temporal features to quantify modality uncertainty arising from input and fluctuation noise. 
The transformer-based multimodal fusion dynamically adjusts each modality's fusion weight based on the quantified uncertainty, mitigating the side effect of low-quality noisy modality on cross-modal relationships.
% Meanwhile, 我们对不确定性表示的学习进行约束，以保证不确定性指标与对预测结果的贡献对应。
% \rev{Meanwhile, we calibrate uncertainty estimation to ensure that the data uncertainty align with its contribution to the health detection accuracy.
\nextrev{Meanwhile, it calibrates modality uncertainty to reflect its contribution to health detection accuracy, ensuring accurate uncertainty estimation to enhance dynamic multimodal fusion.
% 然后，为了从变化数据质量的模态中获得稳定的多模态对齐，我们重建丢失模态通过将不同模态的分布transfer to a commonn space，这有利于提高对其他模态的辅助信息的利用，使得恢复的数据更加准确。（bridges the gap between varying modality-specific feature）
% To achieve stable multimodal alignment within varying data quality, we reconstruct missing modalities by transferring the distributions of different modalities into a common space, allowing effective leverage of complementary information from other modalities for more accurate and reliable data recovery.}
The missing modality reconstruction transfers the modality distributions into a common space, enabling consistent semantic relationships for reliable data recovery.}
% Second, to strengthen critical cross-modal relationship extraction in transformer-based feature fusion, we incorporate adaptive multimodal fusion into the MLLM framework, adaptively adjusting attention scores based on modality contributions to mitigate the inference of low-quality features on cross-modal relationships.
% First, \rev{we minimize the impact of noise on feature extraction by estimating the reliability of each input data. At the same time, we capture the fluctuations and trends of the historical time series, allowing for increased attention to high-variance modalities during periods of significant fluctuations.}
% 最后，我们将动态融合过程运用到多模态大模型框架，根据输入模态特征的质量调节其在注意力计算中的贡献，在低质量数据中关注到关键特征。
% Finally, we integrate the dynamic fusion into the MLLM framework, adjusting each modality's contribution to attention computation according to the quality of its features, enabling the model to focus on key features even with low-quality data inputs.
The key contributions of this paper are summarized as follows: 
\begin{itemize}
% \item We propose a novel outdoor health monitoring framework capable of enabling robust multimodal fusion in complicated noisy environments. 
\item We design a novel modality uncertainty quantification scheme that jointly estimates input and fluctuation noises via current and temporal feature variance, \newrev{allowing the model to distinguish useful health-related variations from irrelevant noise, which is rarely addressed in prior works.}
% to estimate modality uncertainty due to input and fluctuation noise by leveraging current and temporal features.
\item We devise a transformer-based multimodal fusion strategy to \newrev{dynamically adjust and calibrate both modality weights and cross-modal attention}, improving robustness to noisy inputs. \newrev{To our knowledge, this is the first MLLM framework specifically tailored for outdoor health monitoring.}
\item We design a modality reconstruction network to achieve stable multimodal alignment by transferring fluctuating modality distributions into a common space, representing a significantly novel approach.
% \item We propose an adaptive multimodal fusion strategy that quantifies data uncertainty arising from quality degradation and the high-variance fluctuations in health biomarkers, which are used for the dynamic modality-specific weights for multimodal fusion. 
% \item We integrate the adaptive multimodal fusion into the MLLM framework, introducing a transformer-based confidence fusion method to adjust the attention confidence for each modality, which enhances the focus on critical cross-modal relationship even with the presence of low-quality features.
\item We empirically evaluate \name with extensive experiments. The results demonstrate that our scheme outperforms the state-of-the-art frameworks in detection accuracy and the effectiveness of each well-designed component in \name.
\end{itemize}

The rest of this paper is organized as follows. 
% Sec.~\ref{sec:motivation} introduces the motivation for \name by highlighting the limitations of existing health monitoring systems.
Sec.~\ref{sec:related_work} discusses related work and technical limitations. 
Sec.~\ref{sec:system_design} presents the system design of \name. Sec.~\ref{sec:implementation} describes system experimental setup, followed by the performance evaluation in Section~\ref{sec:evaluation}. % Section~\ref{sec:related_work} discusses related work and technical limitations. 
Finally, conclusions are outlined in Sec.~\ref{sec:conclusion}.

% % 9. organization of this paper
% \rev{The rest of this paper is organized as follows. 
% Sec.~\ref{sec:related_work} discusses related work and technical limitations. 
% Sec.~\ref{sec:system_design} first introduces the research problem and then presents the system design of \name.
% Sec.~\ref{sec:evaluation} describes experimental setup and performance evaluation. Finally, conclusions and future research directions are outlined in Sec.~\ref{sec:conclusion}.}

% \vspace{-0.1cm}
\section{Related Work} \label{sec:related_work}
{\bf{MLLMs for health monitoring: }}
Transformer-based language models have achieved remarkable success, paving the way for the development of even larger and more powerful models, such as GPT-4~\cite{bubeck2023sparks}, FLAN-T5~\cite{chung2024scaling}, and LLaMA~\cite{touvron2023llama}. 
% These models demonstrate promising performance across diverse tasks, driving significant advancements in training specialized language models for medical domain. 
The integration of LLMs in healthcare has emerged as a rapidly growing field, with models like BioMedLM~\cite{bolton2022biomedlm}, BioGPT~\cite{luo2022biogpt}, and Med-PaLM~\cite{singhal2023large} fine-tuned on medical data, achieving notable results on biomedical benchmarks and demonstrating their potential in healthcare applications.
Building on this success, there has been a growing interest in extending LLM capabilities to multimodal perception, including the incorporation of medical images~\cite{li2024llava, moor2023med}, \needrev{audio signals~\cite{nagrani2021attention}}, or wearable sensor data~\cite{kim2024health, liu2023large} to support various mental health and disease detection tasks. 
% To achieve effective cross-modal alignment, various studies leverage techniques such as projection layers~\cite{}, intermediate networks~\cite{}, and feature concatenation~\cite{} within transformer architectures to align multimodal features and learn robust cross-modal representations.
\newrev{However, most existing MLLM approaches focus on controlled environments such as driver monitoring and indoor clinical care, where the change of modality quality is relatively stable. These works often overlook the impact of low-quality modalities on critical feature extraction in multimodal fusion, leading to disproportionate attention on noisy features and misalignment of cross-modal relationships, thereby significantly compromising the accuracy of the health status identification. 
% This limitation results in disproportionate attention being assigned to noisy features, leading to improper misalignment to focus on irrelevant cross-modal relationships, thereby significantly compromising the accuracy of the health status identification.
Despite the growing importance of robust multimodal systems, health monitoring in dynamic and noisy outdoor environments remains largely unexplored in existing literature.}

{\bf{Uncertainty Quantification for Multimodal Fusion: }}
Recent advancements in uncertainty modeling have introduced probabilistic distributions to replace point representations.
A widely adopted framework for uncertainty quantification is the Bayesian deep learning network~\cite{blundell2015weight, gal2016dropout, maddox2019simple}, which models network parameters as probabilistic distributions and learns a posterior distribution based on the training data. 
% Due to the computational intractability of exact posterior inference, various methods~\cite{gal2016dropout, kendall2017uncertainties, maddox2019simple} provide an efficient approximation for uncertainty estimation without requiring significant model modifications. 
\rev{Building on these foundations}, recent works~\cite{subedar2019uncertainty, ji2023map, gao2024embracing} explicitly quantify unimodal uncertainty and adaptively adjust fusion weights, enabling more robust multimodal fusion.
% Although these approaches effectively estimate inherent uncertainty arising from task-related randomness, they often fail to account for the uncertainty introduced by low-quality or noisy inputs.
To mitigate the impact of low-quality or noisy inputs, uncertainty quantification has been incorporated into deep learning models, achieving success in domains like face recognition~\cite{chang2020data}, \needrev{medical image analysis~\cite{kwon2020uncertainty}, and emotion recognition~\cite{harper2020bayesian}}.
\newrev{While prior works have explored uncertainty-aware multimodal fusion, they typically focus on single-type noise (e.g., from missing or degraded inputs) and fail to  distinguish between input noise from environmental disturbances and fluctuation noise from abrupt biomarker changes.
In outdoor monitoring, these two types of uncertainty often co-occur and interact, making it hard for traditional fusion strategies to preserve useful health variations while suppressing irrelevant disturbances.}
% However, most of these methods focus solely on modeling uncertainty in individual inputs, 
% \waitrev{As these fluctuation noises often co-occur and interact with input noises, their intertwined effects complicate modality uncertainty quantification to preserve useful health-related variations while suppressing irrelevant disturbances.  misinterpret early biomarkers as irrelevant disturbances leads to increased false alarms or missed detections of abnormal health status.}
Their similarity may cause models to misinterpret early biomarkers as noise, leading to increased false alarms or missed detections of abnormal health status.
% Deep learning models are prone to overconfident predictions, which can lead to unreliable uncertainty estimates and suboptimal decision-making. Consequently, uncertainty calibration has received considerable attention in recent years to improve the reliability of these estimates. For example, some approaches enhance uncertainty calibration by maximizing the correlation between prediction accuracy and uncertainty scores, thereby improving model confidence.
\needrev{Furthermore, although uncertainty calibration~\cite{guo2017calibration, kuleshov2018accurate, mukhoti2020calibrating} has gained increasing attention for mitigating unreliable uncertainty estimates and suboptimal decisions, most existing methods focus on calibrating each modality independently, neglecting the relative uncertainty levels (i.e., data quality levels) across different modalities. Accurately capturing this relative ranking is crucial to calculate modality-specific fusion weights for more reliable multimodal fusion, which has not been well investigated as yet.} % yet remains underexplored
% Our work addresses this gap by explicitly calibrating modality uncertainty within the attention computation, thereby enhancing the alignment between modality confidence and its contribution to health detection accuracy.

{\bf{Modality reconstruction: }}
\rev{Dynamic outdoor environments may cause modality data missing, to mitigate the performance degradation from such missing data, extensive researches develop two data recovery strategies: learning joint multimodal representations~\cite{wang2020transmodality, wang2020multimodal} and generating missing data from available modalities~\cite{zhao2021missing, zhou2021memorizing, perera2019ocgan}.}
Joint multimodal representation learning focuses on capturing shared semantic information to enable robust cross-modal feature extraction under incomplete inputs~\cite{wang2020multimodal, wang2020transmodality}.
For instance, TransModality~\cite{wang2020transmodality} adopts a transformer-based architecture to align features across modalities using inter-modality correlations, thereby mitigating the performance degradation when inputs are incomplete.
\waitrev{In contrast, generative methods, such as AutoEncoders and Variational AutoEncoders, aims to explicitly reconstruct missing modalities by learning shared semantic features from various modalities and decoding them to recover absent information~\cite{zhao2021missing, zhou2021memorizing, perera2019ocgan}. 
For example, MMIN~\cite{zhao2021missing} encodes multimodal inputs into a shared latent space and enforces semantic consistency to directly ``imagine" missing modalities from the available inputs.}
However, these models primarily focus on learning stable correlations between modalities under the assumption that all modalities are of high quality and reliable, overlooking the fluctuating distributions of individual modalities caused by \rev{variations in data quality}. This fluctuating modality distribution disrupt stable cross-modal alignment and hinder consistent correlation learning, potentially leading to inaccurate data recovery that fail to capture critical details.
\section{System Design} \label{sec:system_design}
\newrev{In this section, we introduce \name, the first dynamic uncertainty-aware multimodal fusion framework tailored to outdoor health monitoring where noise is more complex and dynamic, as two distinct but co-occurring sources of uncertainty significantly affect detection accuracy.}
% an adaptive multimodal fusion system to detect critical health status under varying data quality. 
% \rev{As a result, all these issues underscore the necessity} to precisely estimate the reliability and sensitivity of each multimodal sample to achieve adaptive transformer-based feature fusion in dynamic changing driving environments.
Our key idea is \nextrev{to accurately quantify data uncertainty arising from input noise and fluctuation noise and design dynamic weights for transformer-based multimodal fusion in MLLMs, while recovering missing data through other noisy modalities within a common feature space}. 
In what follows, we first outline the system overview and training procedure, and then present a detailed description of the system architecture.

\begin{figure*}[t] 
\centering
\includegraphics[width=0.87\linewidth]{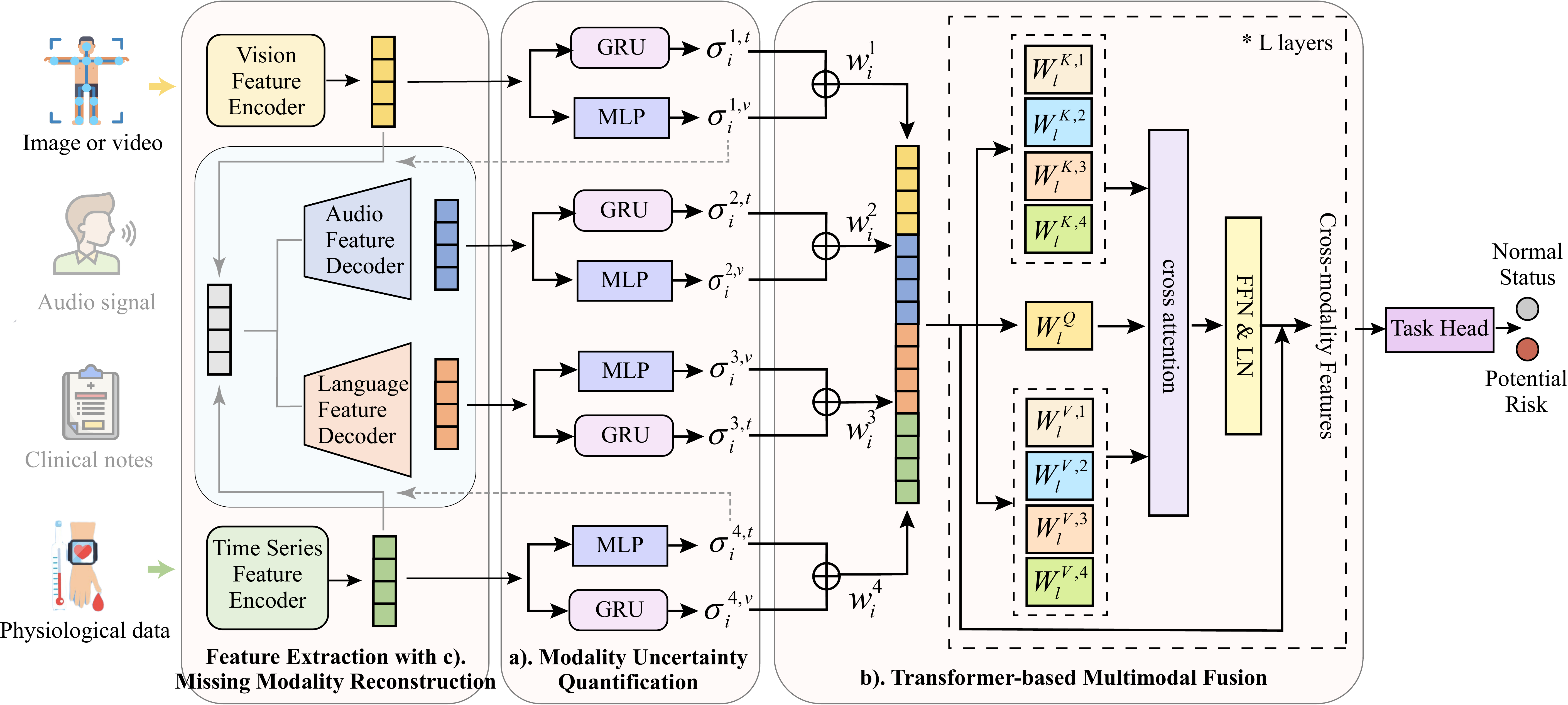}
\caption{The framework of the proposed \name, which consists of three crucial modules: a). modality uncertainty quantification, b). transformer-based multimodal fusion, and c). missing modality reconstruction.}
\label{fig:overview}
% \vspace{-2ex}
\end{figure*}

\vspace{-0.15cm}
\subsection{Overview}
Our design comprises three key components: modality uncertainty quantification, transformer-based multimodal fusion, and missing modality reconstruction. 
% 考虑 high variance inherent in certain health conditions，同时估计输入数据对预测结果的贡献和状态变化趋势作为不确定度，在状态波动剧烈时增加对高方差模态的关注(Sec.~\ref{sec:fusion_uncertainty})，从而抑制噪声数据导致的特征提取不准确和错误预测分类结果。将数据不确定度作为模态权重，动态融合多模态特征，以关注更有信息量的特征 (Sec.~\ref{sec:fusion_weight})，同时根据约束估计的不确定度与预测结果正确性相关，保证估计的合理性 (Sec.~\ref{sec:fusion_calibration})。
The modality uncertainty quantification module estimates the uncertainty of each multimodal input, accounting for input uncertainty arising from current inputs (Sec. III-C1) and fluctuation uncertainty from temporal features (Sec. III-C2).
% Fusion weights are then dynamically assigned to different inputs within a single multimodal sample, balancing the uncertainty arising from input reliability and fluctuation sensitivity (Sec.~\ref{sec:fusion_weight}). 
% Meanwhile, we calibrate the learning of uncertainty representations to ensure that the data uncertainty align with its contribution to the health detection accuracy (Sec.~\ref{sec:fusion_calibration}).
% 我们也考虑了模态丢失对数据不确定度估计的影响，设计一种利用低质量模态恢复数据的方法，在同一空间对齐多模态特征，以消除模态分布变化对稳定的跨模态关系学习的影响(Sec.~\ref{sec:sec:modality_reconstruction})
% We design a modality reconstruction network to bridge the gap between varying modality-specific feature for stable multimodal alignment by mapping modality distributions into a common space.
% Finally, 为了使多模态融合更好的关注到关键的跨模态关系，we develop the transformer-based confidence fusion module to 输入模态特征的质量调节其在注意力计算中的贡献，在低质量数据中关注到关键特征 (Sec.~\ref{sec:transformer_fusion})。
To achieve uncertainty-aware fusion with low-quality modalities, we develop the transformer-based multimodal fusion module that \newrev{first assigns fusion weights across modalities to suppress unreliable inputs while retaining informative fluctuations (Sec. III-D1) and then dynamically adjusts their \needrev{cross-modal attentions} within the transformer architecture to enable robust and adaptive monitoring (Sec. III-D2).}
% adaptively adjusts attention scores based on the contributions of individual modality features
% Fusion weights are then dynamically assigned to different inputs within a single multimodal sample, balancing the uncertainty arising from input reliability and fluctuation sensitivity (Sec.~\ref{sec:fusion_weight}). 
% The overall system architecture of \name is shown in Figure~\ref{fig:overview}.
Meanwhile, we calibrate \rev{uncertainty representations} to ensure the accurate estimation of modality uncertainty for cross-modal fusion (Sec. III-D3).
% 我们也考虑了模态丢失对数据不确定度估计的影响，设计一种利用低质量模态恢复数据的方法，在同一空间对齐多模态特征，以消除模态分布变化对稳定的跨模态关系学习的影响(Sec.~\ref{sec:modality_reconstruction})
Finally, the missing modality reconstruction module allows stable multimodal alignment by mapping fluctuating modality distributions into a common feature space (Sec.~\ref{sec:modality_reconstruction}).

As shown in Fig.~\ref{fig:overview}, the training procedure of \name follows five steps: i) Based on the features of each modality, the proposed modality uncertainty quantification module estimates the uncertainties of input and fluctuation noise separately, ii) For missing data, the modality reconstruction module recovers the missing features from the available modalities, iii) The recovered and existing modality features are combined into multimodal representations through adaptive modality weight assignment according to the quantified uncertainties, and then iv) These weighted multimodal representations are processed by the dynamic cross-modal fusion module, which dynamically captures cross-modal correlations and predicts health monitoring results, finally v) The uncertainty calibration module optimizes the uncertainty estimations by aligning each modality’s contribution with its detection accuracy.
% i) 每个模态数据通过特征编码器提取特征并根据当前和时域数据分别量化输入噪声和波动噪声的不确定性, ii) 对于丢失的模态，丢失模态重建模块根据可用模态的特征恢复丢失的特征, iii) 不同模态的特征进行自适应的多模态融合，根据量化的不确定性, and then iv) 多模态特征输入transformer，通过动态跨模态融合更有效的提取跨模态关系，并输出健康监测结果, finally v) 不确定性校正模块根据单模态对识别正确率的贡献来优化各个模态不确定性的估计.

% relationship among components, influence of each module
% 设计了xx来达到xx目的，从而实现了更好的效果。

\vspace{-0.1cm}
\subsection{System Model} \label{sec:problem_statement}
\name is designed for monitoring potential health emergencies using multiple sensors, providing timely alerts upon detecting significant changes in health biomarkers.
% Those health issues \needrev{(e.g., heart failure, cardiovascular disease, and strokes)} have distinct behavioral and physiological patterns in the early stage, for example severe chest pain, reduced breathing and rapid heartbeat~\cite{}. Continuous tracking and analysis of an individual’s behavioral and physiological health biomarkers enables early detection of potential health problems, thereby reducing health risks and preventing health deterioration through timely intervention.
% As a result, we utilize \rev{multiple} sensors to capture these early signs in different perspectives and \waitrev{train models for each person}, \rev{expecting} early identification of potential health issues under noisy or missing multimodal data.
By leveraging multimodal data, the system captures both behavioral and physiological health biomarkers (e.g., severe chest pain, reduced breathing, and rapid heartbeat) and trains a fusion model resilient to environmental noise. This enables real-time health status detection, mitigating risks, and preventing health deterioration through early intervention.
% As a result, we utilize multimodal data to capture the behavioral and physiological health biomarkers (e.g, severe chest pain, reduced breathing and rapid heartbeat) and train a fusion model resilient to varying environmental noise, enabling timely health status detection to mitigate risks and prevent health deterioration through early intervention.

% data, feature encoder, fusion, label
We denote the multimodal dataset as $X=\{x_1, x_2, ..., x_N\}$, where $x_i=\{x_i^m, m=1, ..., M\}$ is the $i$-th multimodal subdataset containing sensory data from $M$ modalities and $x^m_i$ represents the $i$-th multimodal data sample corresponding to the $m$-th modality.
% we denote $x^m_i$ as the $m$-th modality in the $i$-th multimodal data sample, where $X=\{x_1, x_2, ..., x_N\}$ contains $N$ multimodal data samples and each sample $x_i=\{x_i^m, m=1, ..., M\}$ consists synchronized sensing data from $M$ modalities. 
The feature representation of the $i$-th multimodal data sample corresponding to the $m$-th modality is represented as $z^m_i = f_{\theta_m}(x^m_i)$, which is extracted by the unimodal feature encoder $f_{\theta_m}(\cdot)$ for $m$-th modality. Therefore, the joint multimodal feature representation is denoted by $f_i=[z^1_i, z^2_i, ..., z^M_i]$. The prediction of health status is denoted by $\hat{y}_i$, which is obtained by feeding $f_i$ into the task head. The model is updated by minimizing the loss function of the predicted status $\hat{y}_i$ and the ground truth label ${y}_i$. 

% 将特征建模从point representations转为probability distribution来估计sample对预测结果的不确定性，并根据数据的不确定性来确定每个输入的融合权重和恢复丢失数据，从而抑制噪声数据导致的特征提取不准确和错误预测分类结果。
% \nextrev{To improve the timeliness and reliability of health status detection with noisy inputs, as Figure~\ref{fig:overview} illustrates, instead of extracting point feature representations $z^m_i$ of each unimodal features, we utilize the probability distribution to estimate the uncertainty of each sample's contribution to the \rev{model predictions} and dynamically assign fusion weights based on this data uncertainty.}

\vspace{-0.1cm}
\subsection{Modality Uncertainty Quantification} \label{sec:dynamic_fusion}
{\textit{1) \textbf{Input uncertainty quantification.}}} \label{sec:fusion_uncertainty}
% 挑战难点，解决方案理论支撑，目标，总结关键部分
% Recalling Sec.~\ref{sec:mtv_fusion}, 动态驾驶环境导致数据质量降低，噪声的引入干扰了模型分辨噪声和有用信息，难以提取到有区分度的特征，从而降低使用多模态数据进行健康状态识别的可靠性.
Dynamic environments lead to varying data uncertainty, where sensors may introduce input noise from data acquisition due to environmental changes.
For instance, sudden shadows, poor lighting, and occlusions lead to low-quality images, while physiological signals may be interfered by vigorous movement or changes in body posture.
These low-quality noisy inputs impede discriminative feature extractions for detecting changes in health biomarkers, disrupting the model's ability to differentiate critical health signals from irrelevant information and thus compromising the reliability for health status recognition.
% 
% \waitrev{fixed fusion weights ...}

% 不确定性建模是一种新兴的方法来评估数据的不确性度，不确定性高的数据说明提取的特征不能准确xxx，反映了噪声对xx的影响，可以通过降低关注来xxx。
The variance of feature distributions allows for the assessment of each input's contribution to the overall \rev{model prediction}, \needrev{showcasing its potential to capture input uncertainty}~\cite{subedar2019uncertainty, ji2023map, gao2024embracing}.
% \rev{Recent advancements in uncertainty modeling~\cite{subedar2019uncertainty, ji2023map, gao2024embracing} estimate data uncertainty through the variance of feature distributions, allowing for the assessment of each input's contribution to the overall \rev{model prediction}.}
% 使用方差作为不确定性的原理，做法
By replacing point feature representations with probabilistic distributions, the variance of feature distribution quantifies the dispersion of data around its stable point representations. 
A high variance in feature representation indicates inconsistent model responses to similar inputs, revealing greater ambiguity in feature extraction from varying low-quality noisy modalities. As a result, a larger variance reflects higher uncertainty or unpredictability in the feature representations.
% directly influencing the reliability of data to the \rev{model predictions} for a given input.
% Consequently, low variance indicates that the extracted features are closely clustered around stable patterns, reflecting a low level of unpredictability or uncertainty in the feature representation. 
% \needrev{This motivate us to prioritize data with low uncertainty in multimodal feature fusion to mitigate the negative impact of noise on model performance.}
\rev{Therefore, to quantify the input uncertainty}, the low-quality noisy data is represented with a probabilistic distribution, with feature variance serving as a measure of the input uncertainty. 

Specifically, after extracting features from the feature encoder of the $m$-th modality, we model the deterministic feature representation $z^m_i$ as a multivariate Gaussian distribution $N(\mu^{m,v}_i, {\Sigma^{m,v}_i})$~\cite{ji2023map, sanchez2021affective}, where $\mu^{m,v}_i$ represents the mean of the features and ${\Sigma^{m,v}_i}$ denotes the feature variance from noisy input data.
\begin{align} \label{eq:uncertainty_modeling}
p(z^m_i|x^m_i) \sim ~&N(\mu^{m,v}_i, {\Sigma^{m,v}_i}), \\
\mu^{m,v}_i = f_{\varphi^m}(x^m_i)&,~ \Sigma^{m,v}_i = f_{\psi^m}(x^m_i),
\end{align}
where $f_{\varphi^m}(\cdot)$ and $f_{\psi^m}(\cdot)$ represent two deep learning networks to estimate mean $\mu^{m,v}_i$ and variance $\Sigma^{m,v}_i$, respectively.

The norm value $|| \Sigma^{m,v}_i ||_2$ aggregates the variances across all feature dimensions, \rev{reflecting the uncertainty level of the $m$-th modality for health status classification under varying input noise in dynamic environments.}
Therefore, after normalization of the variance norm values, the input uncertainty $r^m_i$ of the $m$-th modality in the $i$-th sample is denoted as
\begin{equation} \label{eq:data_uncertanty}
% r^m_i = \frac{1/|| \sigma^{m,v}_i ||^2_2}{\sum_m^M {1/|| \sigma^{m,v}_i ||^2_2}},~ s^m_i = \frac{1/|| \sigma^{m,t}_i ||^2_2}{\sum_m^M {1/|| \sigma^{m,t}_i ||^2_2}}
r^m_i = || \Sigma^{m,v}_i ||_2.
% s^m_i = || \sigma^{m,t}_i ||^2_2.
\end{equation}

{\textit{2) \textbf{Fluctuation uncertainty quantification.}}} \label{sec:fusion_contribution}
% 方差可以表示当前输入的不确定性  variance - uncertainty - noise impact and data contribution
By modeling modality features as probabilistic distributions in Eqn.~\eqref{eq:uncertainty_modeling}, existing methods~\cite{subedar2019uncertainty, ji2023map, gao2024embracing, chang2020data, kwon2020uncertainty} utilize feature variance to represent the uncertainty of input noise on data contributions to health status identification.
% 动态驾驶环境的特点，挑战
% % However, the driving environment and emergency maneuvers influence the physiological and behavioral signals being monitored and induce high-variance fluctuations in health biomarkers.
% However, outdoor environments induce significant fluctuation in health biomarkers just as \needrev{sudden traffic accidents} may trigger the abrupt changes of physiological signals and behaviors.
\newrev{However, they typically account for only single-source uncertainty, overlooking two intertwined uncertainties in outdoor monitoring: input noise from environmental disturbances and fluctuation noise from rapid physiological changes.}
\waitrev{In practice, abrupt dynamic environmental changes, such as sudden traffic accidents, can trigger cardiovascular emergencies, which presents early biomarkers such as physiological signal fluctuations before health deterioration.}
% 在由急刹车等导致的高波动性状态的健康监测中，缺点，影响
% Failing to account for these interference-sensitive fluctuations in uncertainty modeling causes the model to misinterpret these early biomarkers as noise, potentially leading to delayed health status detection or even missing important events.
Without accounting for these fluctuation-sensitive patterns, models may misclassify early biomarkers as noise, leading to delayed or missed detection of critical health events.
% These limitations force traditional algorithms to deliver only comparable or even inferior performance in dynamic driving environments.

% To combat these uncertainties, we model input uncertainty from environmental noise and sensor degradation, and fluctuation uncertainty from physiological dynamics. As they stem from distinct sources, we treat them separately to capture their unique characteristics, enhancing fusion robustness and sensitivity to early biomarkers.
\newrev{To address the intertwined uncertainties in outdoor monitoring, we separately model input uncertainty from environmental noise and sensor degradation, and fluctuation uncertainty from physiological dynamics, capturing their distinct characteristics to improve fusion robustness and sensitivity to early biomarkers.
% it alone cannot fully capture complex noise structures. To compensate for this limitation
While variance serves as a core indicator for uncertainty quantification, it captures only the magnitude of fluctuations. To model temporal dynamics and fluctuation patterns, we introduce temporal modeling of physiological signals. 
% This combination allows the model to better handle complex noise and improve the accuracy and reliability of health status detection.
This combination allows the model to prioritize reliable features while maintaining focus on critical health signals under changing environments.}
% By balancing the influence of input and fluctuation uncertainties, the model can prioritize reliable features while maintaining focus on critical health signals under changing environments, enhancing fusion robustness and sensitivity to early biomarkers.}

The feature variance of a single input data only reflects the reliability of that specific input, whereas extracting features from a time series captures the structural patterns and changes of data contribution. 
% Time-series feature variance represents the fluctuations and trend variability within a given period. 
% A smaller time-series feature variance indicates stable feature distributions within a given period, supporting the consideration of that health biomarker to maintain sensitivity for timely detection to these critical changes. When abrupt fluctuations occur in health biomarkers, the high value of input uncertainty and the stability of historical temporal data suggests that these variations may result from \needrev{sudden emergencies}. Persistent high fluctuations over time indicate significant modality uncertainty, warranting a reduced focus on this modality.
\waitrev{Low feature variance in time-series indicates a stable trend over a given period, providing a reliable \rev{base level} for normal status. When abrupt fluctuations occur in health biomarkers, the stability of historical temporal data suggests that these changes are more likely to be early health biomarkers rather than noise, enabling the model to maintain sensitivity to these critical changes. However, persistent high fluctuations over time indicate significant interference, leading to increased uncertainty which warrant a reduced reliance on the affected modality to minimize misclassifications and false alarms.}
Building on this insight, we model temporal dynamics as probabilistic distributions to better capture health-related fluctuations and enhance the timely detection of abnormal health status.
% \newrev{Separating input and fluctuation noises ensures a comprehensive evaluation of uncertainty, capturing both external (e.g., input noise) and intrinsic (e.g., sensitivity of health biomarkers) factors that influence model performance.}

For the current input $x^m_i$, we leverage $T$ historical temporal features $[z^m_i, z^m_{i-1}, \dots, z^m_{i-T}]$ to capture dynamic relationships based on a temporal network like GRU.
% \rev{where $T$ is the number of historical samples in the input sequence.} 
% Here the GRU extracts temporal features that encapsulate the dynamic relationships and trends within the input sequence.  
Then, we learn the probabilistic distribution of temporal features as a multivariate normal distribution $N(\mu^{m,t}_i, {\Sigma^{m,t}_i})$, and capture the time-series feature variance ${\Sigma^{m,t}_i}$ to quantify the fluctuation uncertainty.
The norm of the time-series feature variance, $|| \Sigma^{m,t}_i ||_2$, estimates the \rev{average dispersions} in health biomarkers to detecting critical changes of health status. 
% Conversely, its inverse quantifies 
The fluctuation uncertainty $s^m_i$ for the $i$-th sample can be expressed as
\begin{equation}
    s^m_i = || \Sigma^{m,t}_i ||_2.
\end{equation}
% 虽然我们以方差作为核心不确定性指标，但这只是整体不确定性建模的一部分。为弥补方差在捕捉复杂噪声结构上的不足，我们设计了专门的不确定性机制，通过 GRU 捕捉生理信号的时序动态和波动趋势。方差反映整体波动幅度，时序建模揭示噪声的结构特征和动态变化。两者配合使用，使模型能更好地处理复杂噪声。

% 应该用大模型？？
\begin{figure}[t]
\centering
\subfloat[Modality weights \label{fig:fusion_weight_attn}]{
    \includegraphics[width=0.49\linewidth]{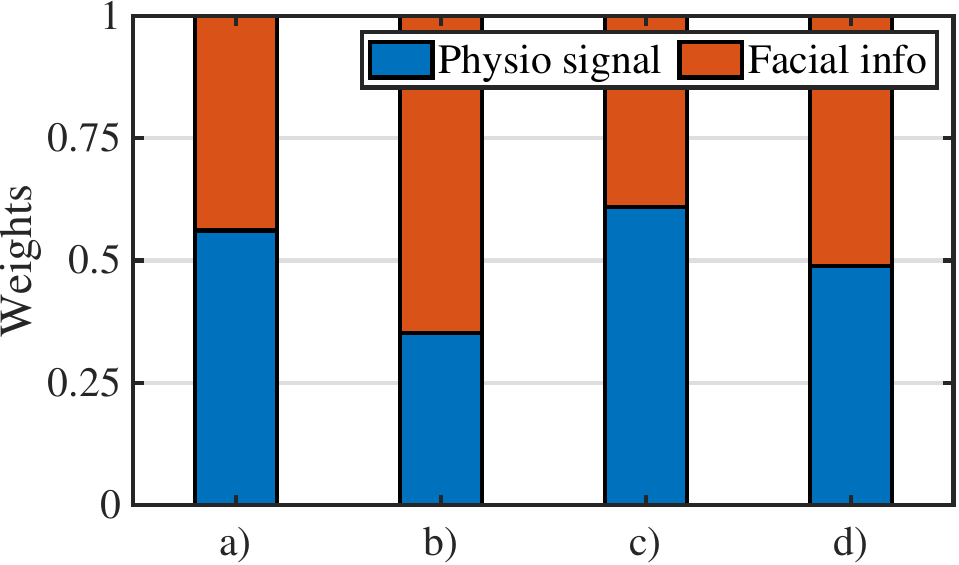}
}
 \subfloat[Detection accuracy \label{fig:fusion_weight_acc}]{
    \includegraphics[width=0.49\linewidth]{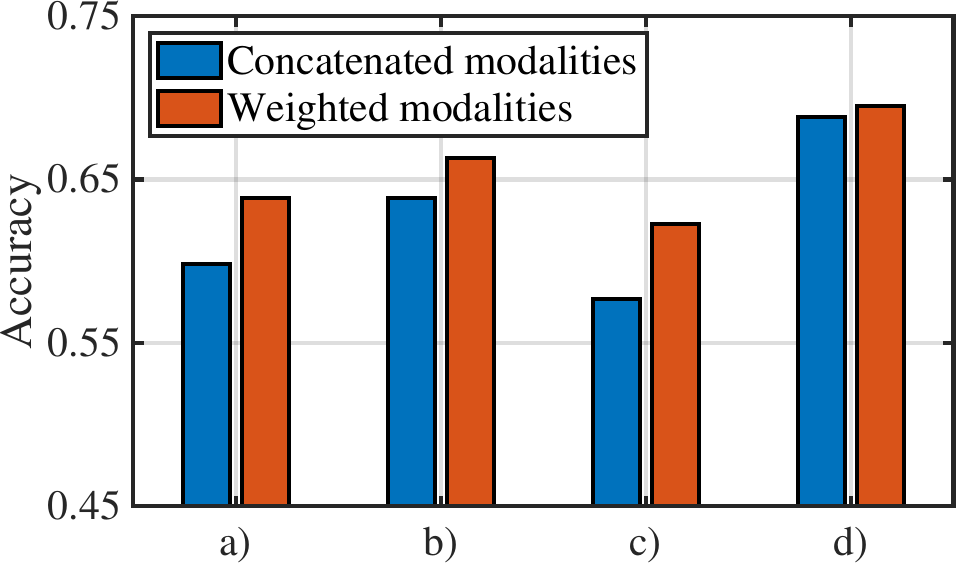}
}
    \caption{Modality contribution to health status recognition under diverse environments \needrev{with a). poor lighting condition; b). body posture changes; c). occlusion; d). normal condition with high-quality data.}}
    \label{fig:fusion_weight}
    \vspace{-2ex}
\end{figure}

\subsection{Transformer-based Multimodal Fusion} \label{sec:transformer_fusion}
{\textit{1) \textbf{Adaptive modality weight assignment.}}} 
\label{sec:fusion_weight}
As input uncertainty and fluctuation uncertainty dynamically change with each input in the multimodal samples, the absence of an adaptive strategy hinders multimodal fusion performance by failing to address the varying impact of low-quality data on detection accuracy. 
To better understand the contribution of each modality to health status detection in diverse environments, we employ a CNN-based multimodal model, \needrev{DeepSense~\cite{yao2017deepsense}}, with an attention module to learn the weights of different modalities on a public multimodal dataset, Stressors~\cite{taamneh2017multimodal}, \nextrev{where the training and testing data from a specific modality is augmented with random noise to simulate the corresponding environments.} 
% 在多模态数据中，不同模态或输入的质量和重要性可能存在显著差异。不同模态在不同环境中对预测结果的贡献不一样。
As shown in Fig.~\ref{fig:fusion_weight_attn}, modality weights vary across different environments, indicating the dynamic contributions of diverse modalities.
\nextrev{Compared with feature concatenation for multimodal fusion}, Fig.~\ref{fig:fusion_weight_acc} shows that prioritizing modalities with greater contributions improves the detection accuracy under low-quality data.
% \needrev{highlighting the necessity to design dynamic cross-modal fusion weights based on modality contributions.}
\newrev{Extensive studies~\cite{subedar2019uncertainty, ji2023map, gao2024embracing} treat uncertainty as a standard way to improve model performance. Following this, we quantify modality uncertainty with the estimation of both input and fluctuation uncertainty and use it to devise dynamic fusion weights}, as shown in Fig.~\ref{fig:overview}. This enables the model to adaptively adjust the contribution of each input during multimodal fusion, ensuring reliable and timely detection in changing environments. 
% This ensures that the predictions are primarily driven by reliable and discriminative data.
% ensures that the model leverages reliable features effectively while mitigating the influence of uncertain or noisy data.

\begin{figure*}[t] 
\centering
\includegraphics[width=\linewidth]{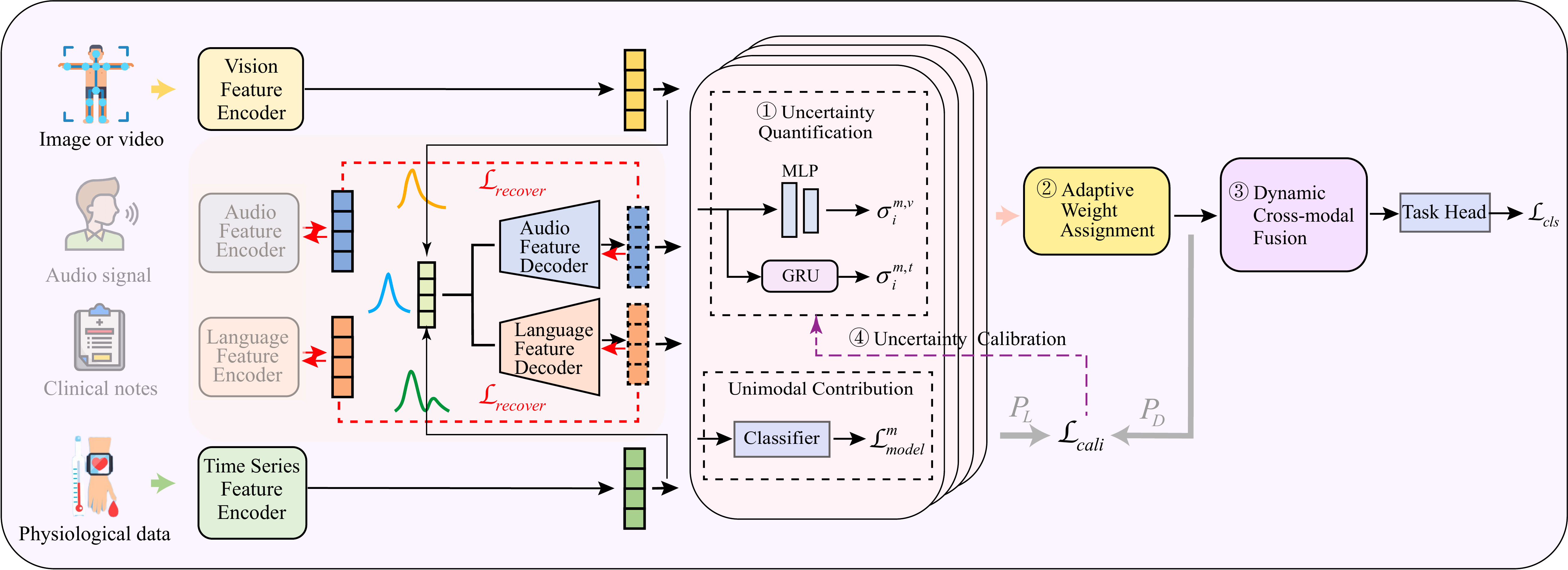}
\caption{The training process of \name with modality uncertainty calibration.}
\label{fig:training}
\vspace{-2ex}
\end{figure*}

For a multimodal sample $x_i=\{x_i^m, m=1, ..., M\}$ with $M$ modalities, we estimate the feature variance
\begin{math} {\Sigma^{m,v}_i} \end{math} and the time-series feature variance \begin{math} {\Sigma^{m,t}_i} \end{math} for each modality.
\nextrev{To accurately detect abnormal health status and dynamically adapt to varying environments,
we compute overall modality uncertainty using the product of $r^m_i$ and $s^m_i$,} \rev{reducing sensitivity to the variance under stable health conditions while increasing focus on sensitive modalities during significant fluctuations.}
By using the inverse of the combined uncertainty as the dynamic weight for each modality in multimodal fusion~\cite{subedar2019uncertainty, ji2023map, gao2024embracing}, the model adaptively prioritizes more reliable and critical modalities in the joint feature representation $f_i$, thereby enhancing the effectiveness of multimodal fusion. The feature fusion for the $i$-th sample can be expressed as \rev{the concatenation of the weighted features from different modalities.} 
\begin{equation} \label{eq:dynamic_fused_feature}
f_i = concat[w^1_i z^1_i, ..., w^M_i z^M_i],~ w^m_i = \frac{1/r^m_i \times 1/s^m_i}{\sum_{j=1}^M (1/r^j_i \times 1/s^j_i)},
\end{equation}
where $w^m_i$ is the dynamic weight for the $m$-th modality in the $i$-th multimodal sample, guiding the model to prioritize cleaner, more informative inputs.

{\textit{2) \textbf{Dynamic cross-modal fusion.}}} \label{sec:transformer_weight}
As explained in Sec.~\ref{sec:introduction}, MLLMs emerge as a promising solution for health monitoring with limited labeled data, leveraging general medical knowledge for fast task adaptation and extensive parameters to detect subtle health changes.
However, current MLLMs~\cite{chan2024medtsllm, li2024llava, moor2023med, kim2024health, liu2023large} often fail to account for the varying quality of different modalities during cross-modal interactions. 
\waitrev{This limitation results in disproportionate attention being assigned to noisy features, leading to improper misalignment to focus on irrelevant cross-modal relationships, thereby significantly compromising the accuracy of the health status identification.}

\newrev{While adaptive multimodal weight assignment proposed in Sec.~\ref{sec:fusion_weight} filters noisy inputs by selecting modalities with lower uncertainty, it focuses on modality-level selection to build a stable fused representation. However, \rev{effectively} extracting \needrev{robust} cross-modal relationships remains a challenge.}
% Conventional self-attention mechanisms compute attention solely based on feature correlations, without considering the varying reliability of different modalities. Consequently, noise-induced distortions can mislead attention allocation, causing the model to focus on irrelevant features and impairing its ability to detect subtle changes in health biomarkers.
Standard self-attention mechanisms compute attention weights based on cross-modal feature correlations without explicitly considering the reliability of each modalities. As a result, noise-induced variations can distort these relationships, and
% \rev{hindering the model's ability to identify the true feature dependencies}. 
% misleading attention scores, resulting in incorrect focus on irrelevant or noisy information.
treating all modalities equally leads to misallocation of attention to irrelevant features, severely impairing the identification of subtle changes in health biomarkers.
\newrev{To overcome this limitation, we incorporate adaptive weighting into the transformer framework. 
During cross-modal fusion, the fusion weights are used to adjust the semantic-level attention across modalities, dynamically controlling each modality’s contribution to the cross-modal semantic fusion and preventing noisy modalities from dominating the joint representation.}
By decoupling attention matrices for different modalities within the shared semantic context, our approach adjusts attention scores based on each modality's dynamic contribution, \nextrev{enabling robust and uncertainty-aware cross-modal feature fusion.}
% better capture cross-modal correlations while minimizing the impact of low-quality modalities.

Using the same attention matrix for all modalities assumes that the transformation works equally well for all data. Therefore, to enhance attention to cross-modal relationships, we decouple the attention matrices for different modalities and adjust their confidence scores based on the uncertainty of modality features, as shown in Fig.~\ref{fig:overview}.
Specifically, the query matrix $Q_{l}$ is computed using a shared projection matrix $W^Q_l$ applied to the output of the previous transformer layer $H_{l-1}$ where the transformer's input is $H_0=f_i$.
% \rev{focusing on capturing shared semantic context.} 
This shared query matrix captures the common semantic context across all modalities, \rev{allowing for the focus on a unified understanding of cross-modal relationships.}
\begin{equation}
    Q_{l} = H_{l-1} W^Q_{l},
\end{equation}

The key and value matrices $K_{l}^i$ and $V_{l}^i$ are generated as a weighted sum across all modalities, where the projection metrics $W^{K,m}_{l}$ and $W^{V,m}_{l}$ are independently learned by each modality, which can be represented as
\begin{align}
    K_{l}^i = \sum_{m=1}^M &w^m_i (W^{K,m}_{l} H_{l-1} \cdot \mathbf{1}(f_i \in z_i^m)), \\
    V_{l}^i = \sum_{m=1}^M &w^m_i (W^{V,m}_{l} H_{l-1} \cdot \mathbf{1}(f_i \in z_i^m)),
\end{align}
\rev{where $\mathbf{1}(f_i \in z_i^m)$ denotes the indicator function, which is 1 if the index of multimodal feature $f_i$ belongs to the $m$-th modality, and 0 otherwise.}
\waitrev{By decoupling attention matrices $W^{K,m}_{l}$ and $W^{V,m}_{l}$, the model isolates the noisy contributions of individual modalities and prevents distortion in cross-modal attention.
% Separate $W^K_m$ and $W^V_m$ ensure that noisy modalities do not distort the representations of other modalities, while
The dynamic weight $w^m_i$ acts as a uncertainty-aware scaling factor to further enhance the confidence of cross-modal feature relationships, while the weighted sum in $K_{l}^i$ and $V_{l}^i$ enables dynamic focus on cross-modal relationships, thereby leading to more accurate detection results}.
% Separate weight matrices $W^K_m$ and $W^V_m$ allow the model to learn modality-specific representations.
% The summation mechanism ensures that all relevant modalities are incorporated into $K$ and $V$ while respecting their individual contributions, preserving complementary information and focusing on the most relevant modalities.

Finally, the cross-modal attention in the $l$-th transformer layer is computed as 
\begin{equation}
    {Attn}^{i}_{l} = Softmax\left(\frac{Q_{l}^i K_{l}^i}{\sqrt{d_K}}\right)V_{l}^i,
\end{equation}
where $d_K$ refers to the dimensionality of the key matrices $K_{l}^i$. The output of one transformer layer $H^{i}_{l}$ is calculated by applying a feedforward network (FFN) and layer normalization (LN) to the cross-modal attention ${Attn}^{i}_{l}$ as $H^{i}_{l} = LN(FFN({Attn}^{i}_{l}))$.
% \needrev{put it together}
% The pseudocode of \name is summarized in Algorithm.~\ref{alg:name_training} in Appendix~\ref{sec:appendix_alg}.

% Specifically, the query matrix $Q$ is computed using a shared projection matrix $W^Q_l$ applied to the representation from previous layer $H_{l-1}$, \rev{focusing on capturing shared semantic context.} 
% The key and value matrices $K$ and $V$ are generated as a weighted sum across all modalities, where each modality \rev{contributes} with its own projection metrics $W^K_m$ and $W^V_m$.
% \rev{Separate $W^K_m$ and $W^V_m$ ensure that noisy modalities do not distort the representations of other modalities. Dynamic weight $w^m_i$ further enhances the confidence of cross-modal feature fusion}, leading to more accurate detection results.
% The self-attention operation is represented as
% \begin{align}
%     &Q = H_{l-1} W^Q, \\
%     K = \sum_m^M &H_{l-1} (w^m_i W^K_m \cdot \mathbf{1}(f_i \in z_i^m), \\
%     V = \sum_m^M &H_{l-1} (w^m_i W^V_m \cdot \mathbf{1}(f_i \in z_i^m), \\
%     Attn &= Softmax(\frac{QK}{\sqrt{d_K}})V
% \end{align}
% where $\mathbf{1}(f_i \in z_i^m)$ denotes the biomarker function for the index of $m$-th modality in the multimodal feature $f_i$ and the transformer's input is $H_0=f_i$.

% 0.75 page
{\textit{3) \textbf{Model training and uncertainty calibration.}}} \label{sec:fusion_calibration}
% 为什么要做
% 我们假设数据不确定性能够准确反映数据对正确预测健康状态的贡献，因此在动态权重分配可以据此对不同模态予以不同权重。
% 不确定性建模为多模态特征融合提供了动态调整权重的依据。但是，准确表示数据对预测健康状态的贡献才能使得融合权重合理分配，避免模型对无用模态的过度依赖，使得模型更依赖于有信息量的模态特征，从而提升多模态健康时间识别的正确性。
Uncertainty quantification measures the contribution of each input, serving as an \rev{evidence} for dynamically adjusting weights in multimodal feature fusion.
However, \nextrev{correctly representing and ranking the relative contribution} of multiple modalities for identifying abnormal health status is crucial to ensure that the fusion weights are assigned appropriately among modalities.
% For example, modality features with lower uncertainty provide more reliable and informative information for accurate predictions. The model prioritizes these features as it is more likely to achieve accurate prediction.
% If the estimated uncertainty does not align with prediction accuracy (e.g., highly uncertain data actually contributes significantly), the model may incorrectly deprioritize reliable inputs, leading to degraded performance.
% 准确表示每个模态数据对预测的真实贡献使得模型对每种模态的依赖更加合理; 通过对模态贡献的相对大小进行排序，模型可以优先聚焦于更有信息量的模态
Accurately representing each modality’s contribution prevents over-dependence on irrelevant features~\cite{nguyen2015deep}, while ranking their relative contributions enables the model to prioritize more reliable and informative modalities~\cite{moon2020confidence}, thereby enhancing the accuracy of multimodal health status recognition.
% Calibration ensures balanced contributions from all modalities
To achieve this, we propose to calibrate the modality uncertainty to align dynamic weights with each modality's contribution to detection accuracy.

\begin{figure}[t]
\centering
\subfloat[Physiological signals \label{fig:calibrate_correlation_phy}]{
    \includegraphics[width=0.49\linewidth]{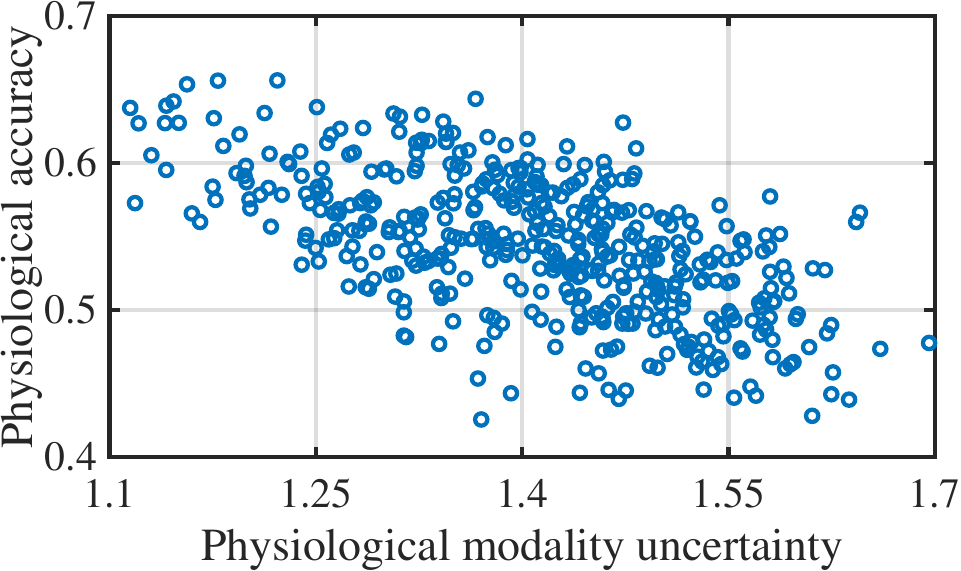}
}
 \subfloat[Visual data \label{fig:calibrate_correlation_visual}]{
    \includegraphics[width=0.49\linewidth]{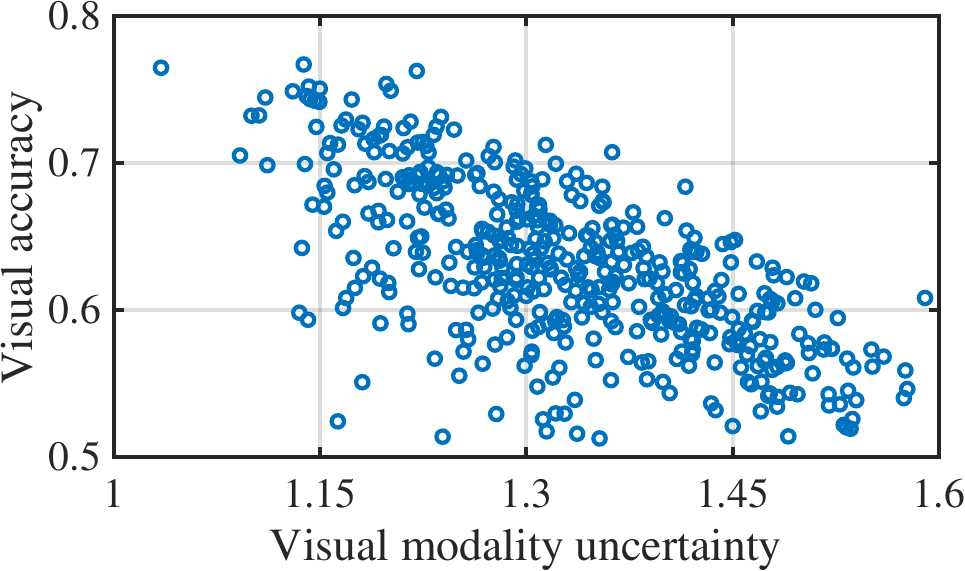}
}
    \caption{The visualized relationship between modality uncertainty and its detection accuracy corresponding to each input in various modalities.}
    \label{fig:calibrate_correlation}
    \vspace{-2ex}
\end{figure}

% 如何做，为什么能做，有什么好处
% 预测误差和数据不确定性呈线性关系，强相关，可用于约束（对齐）方差 linearly correlated with prediction error 
% Since uncertainty reflects the model’s confidence in the input’s reliability and informativeness for prediction, data contribution can be measured by quantifying the uncertainty of a specific input on the model's prediction accuracy.
The contribution for each input directly correlates with its impact on detection accuracy, which implies that higher modality uncertainty should correspond to a greater probability of inaccurate detections. 
% the uncertainty measure need to accurately reflect the likelihood of error in predictions. If a model is highly uncertain, it should correspond to a higher likelihood of prediction errors. 
% Without this calibration, the uncertainty measure might not accurately reflect the likelihood of error in predictions.
% 我们使用DeepSense模型来验证方差和预测误差的关系。其中使用两层全连接层来学习特征分布的方差，对生理信号加入噪声，用特征方差表示数据的不确定性，然后可视化数据方差和预测误差的联合分布。As demonstrated in Fig.~\ref{fig:calibrate_correlation}, 预测误差和数据不确定性呈强相关的线性关系，(经过校正后特征方差和预测误差的线性关系更强)。
% use a two-layer fully connected network to learn the feature variances
To validate the relationship between modality uncertainty and detection accuracy, \needrev{we extract the unimodal features from physiological and visual feature encoders separately} and \needrev{estimate their modality uncertainties}. The joint distribution of \needrev{modality uncertainty} and detection accuracy is visualized in Fig.~\ref{fig:calibrate_correlation}, \rev{which demonstrates a strong linear relationship between modality uncertainty and detection accuracy across different modalities.}
% 因此我们使用识别正确率对数据不确定性进行校正。我们首先估计单模态特征对模型预测的贡献，使用单模态预测误差对每个特征的不确定性进行校正，使每个输入的不确定性与预测误差对应，同时能够有效表达不同模态的相对贡献。
% Thus, we calibrate data uncertainty using detection accuracy. 
Therefore, we estimate the contribution of unimodal features to its detection accuracy as a constraint for calibrating each input's uncertainty. \nextrev{By minimizing the mismatch between the distribution of modality uncertainty and detection accuracy, each input's modality uncertainty is better aligned with its detection accuracy, effectively reflecting the relative contributions across modalities.}

% 怎么做 - Aligning Uncertainty with Prediction Accuracy
% 为了获得每个模态单独对应的模型预测，我们使用分离的特征编码器独立提取每个模态的特征，然后在单模态特征后添加分类器 \phi_m for unimodal predictions, as shown in Fig.~\ref{fig:dynamic_fusion}. 通过交叉熵函数根据ground truth y_i 来计算每个输出的识别正确率。单模态模型训练的损失函数为
To obtain the detection accuracy corresponding to each input, we use unimodal feature encoders to extract features from each modality independently, and train a classifier $\phi_m(\cdot)$ for \rev{model predictions without multimodal fusion}, \needrev{as shown in Fig.~\ref{fig:training}}. \rev{The unimodal detection accuracy of the $i$-th sample is calculated using the cross-entropy (CE) loss function with the ground truth label $y_i$}, which is expressed as 
\begin{equation}
\mathcal{L}_{modal}^m = CE(\phi_m(z^{m}_i),y_i), 
% \mathcal{L}_{cls} = \frac{1}{K+1} (CE(\phi(\mu^{m,v}_i),y_i) + \sum_k^K CE((\phi(z^{m}_{i+k}),y_i))
\end{equation}

% 我们收集每个模态输入的数据不确定性和与之对应的模型预测正确率，并将其组成一个向量。
Then we combine the modality uncertainty and its corresponding detection accuracy from each input into two metrics as $P_D=\left[ w^1_i, w^2_i, ..., w^M_i \right]$ and $P_L=\left[ \mathcal{L}_{modal}^1, \mathcal{L}_{modal}^2, ..., \mathcal{L}_{modal}^M \right]$.
% \begin{align}
% P_D &= \left[ w^1_i, w^2_i, ..., w^M_i \right], \\
% P_L &= \left[ CE(\phi(z^{1}_i),y_i), CE(\phi(z^{2}_i),y_i), ..., CE(\phi(z^{M}_i),y_i) \right]
% \end{align}

% 我们对数据不确定性的向量进行校正，使其与模型预测的向量一致，由此来同时校正和排序不同模态的不确定性。
% 通过最小化mismatch between distance distribution and variance-norm distribution来对齐learnable parameters和预测误差，使用Jensen-Shannon divergence 使得两个向量的分布逼近.
To both calibrate and rank the uncertainty of different modalities, we minimize the mismatch between the distribution of modality uncertainty and unimodal detection accuracy. \needrev{Jensen-Shannon divergence, as a symmetric measure of similarity between probability distributions, ensures a balanced alignment between modality uncertainty and detection accuracy.} Therefore, we use the Jensen-Shannon divergence to approximate the distribution of the two metrics $P_D$ and $P_L$, which is given by
% we calibrate the data uncertainty vector to align with the model's prediction vector. This is achieved by minimizing the mismatch
\begin{equation}
\mathcal{L}_{cali} = \frac{1}{2} \left( KL(P_D||P_L) + KL(P_L||P_D) \right)
\end{equation}

% 根据估计的不确定性进行动态权重分配，\name的最终目标是提升多模态融合后的识别正确率。我们通过xxx来实现训练目标\mathcal{L}_{cls}，其中\phi对多模态特征进行分类。
% \nextrev{The training objective of \name is to classify different health events with the fused multimodal features as $\mathcal{L}_{cls}=CE(\phi(f_i),y_i)$, where $\phi$ is the classification head that identifies health events from multimodal features $f_i$ in Eqn.~\eqref{eq:dynamic_fused_feature}.}
The training objective of \name is to classify diverse health status by leveraging fused multimodal features. To achieve this, the \rev{detection accuracy is optimized} by the cross-entropy loss function, which measures the discrepancy between the predictions after transformer-based multimodal fusion and the ground truth labels. The loss function for model training is formulated as
\begin{equation}
\mathcal{L}_{cls} = CE(\phi(H^i_L),y_i),
\end{equation}
where $\phi(\cdot)$ is the task head that classifies health status from cross-modal features $H^i_L$ after $L$ transformer layers.

To the end, combining the transformer-based multimodal fusion for dynamic weight assignment and the calibration of modality uncertainty to unimodal detection accuracy, the training loss for dynamic multimodal fusion is defined as
\begin{equation}
\mathcal{L}_{dyn} = \sum_{i=1}^N (\mathcal{L}_{cls} + \lambda_a \sum_{m=1}^M \mathcal{L}_{modal}^m + \lambda_c \mathcal{L}_{cali})
\end{equation}
\waitrev{where $\lambda_a$ is the balance weight for unimodal classification training, and $\lambda_c$ controls the regularization strength of modality uncertainty calibration.}

% 0.5 + 0.25 page
\vspace{-0.1cm}
\subsection{Missing Modality Reconstruction} \label{sec:modality_reconstruction}

\begin{figure}[t]
\centering
\subfloat[Feature distribution \label{fig:missing_distribution}]{
    \includegraphics[width=0.49\linewidth]{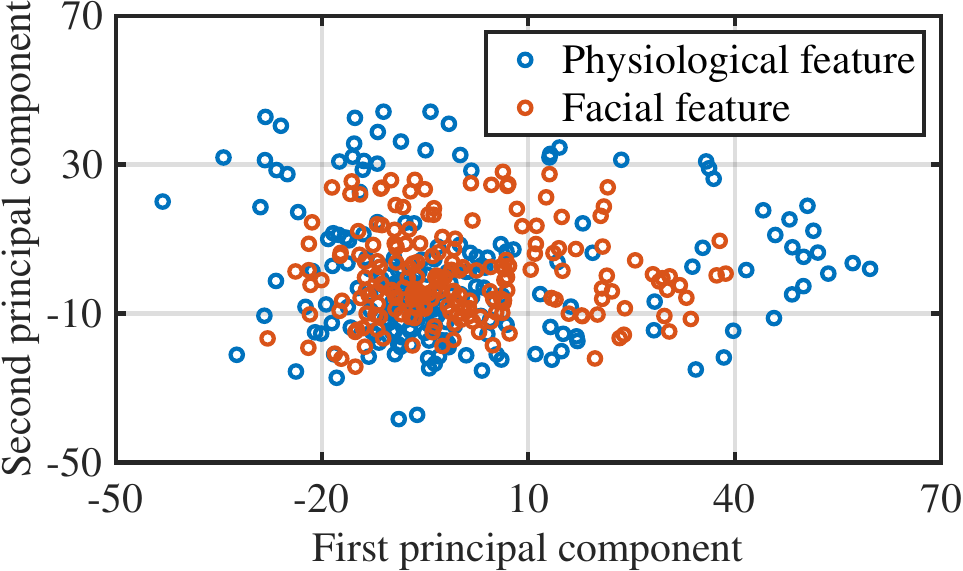}
}
\subfloat[Detection with recovered data \label{fig:missing_predict}]{
    \includegraphics[width=0.49\linewidth]{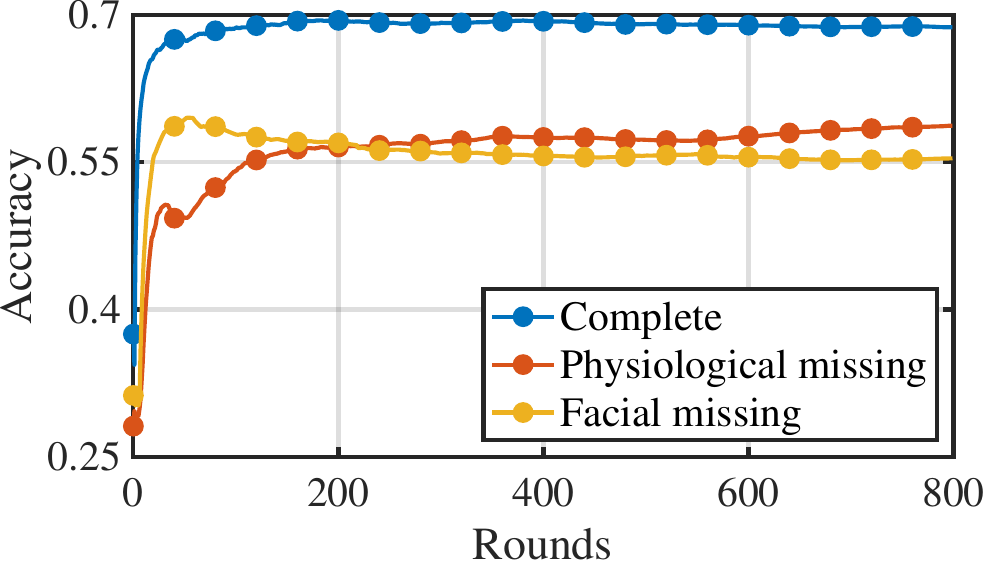}
}
    \caption{\needrev{Impact of distribution variability and the detection accuracy with the missing heartbeat/image recovery.}}
    \label{fig:missing_impute}
    \vspace{-2ex}
\end{figure}

% 现有模态补偿方案忽略模态分布不一致的问题，modality-specific features不能很好的转换, leading to inaccurate recovered data；同时噪声数据的补偿效果不好 -> 我们提出一种噪声数据的模态补偿方法，首先考虑不同模态数据的分布不一致，先将模态分布转为统一的，然后通过方差加权置信度，降低噪声影响 -> \mathcal{L}_{rec} = || \hat_z - z || ^2 -> 模态丢失时的动态融合过程

% 由于动态驾驶环境，模态丢失在车内监测中很常见，we propose a modality reconstruction network。多模态数据reveal相似的语义，因此可以通过学习模态之间的映射关系，利用现有模态来恢复丢失数据。
% 然而，现有的数据恢复方案通常假设高质量的训练数据，
As dynamic outdoor environments also cause modality missing, we propose a modality reconstruction network to recover the feature representations of the missing modality before transformer-based multimodal fusion.
Multimodal data reveals similar semantics from diverse perspectives, allowing missing data to be recovered by leveraging cross-modal correlations learned from the available modalities.
However, current data recovery methods~\cite{wang2020transmodality, wang2020multimodal, zhao2021missing, zhou2021memorizing, perera2019ocgan} \waitrev{rely on stable correlations from remaining modalities}, failing to account for the fluctuating modality distributions caused by the varying data quality of other modalities.
Fluctuating distributions exacerbate the discrepancy between modalities, \nextrev{complicating cross-modal alignment to capture consistent semantic relationships.}
% introducing instability in cross-modal relationship learning.
% making it harder for models to leverage complementary information effectively.
To motivate our design of modality reconstruction module, we apply PCA to visualize the first two components of feature distributions for heartbeat and visual samples in the Stressors dataset.
Then, we deploy MMIN~\cite{zhao2021missing}, a data recovery network from multiple modalities, to reconstruct missing heartbeat or visual data when the other modality is available \rev{under varying environments}. \needrev{The detection accuracy is then evaluated with the multimodal data with/without missing.} 
As shown in Fig.~\ref{fig:missing_distribution}, the \rev{changing environments} result in significant discrepancies in the feature distributions across modalities, causing cross-modal misalignment.
Consequently, Fig.\ref{fig:missing_predict} illustrates a substantial gap in detection accuracy \rev{between recovered and complete data.}
% As mentioned in Fig.~\ref{fig:missing_impute}, such changing modality distributions hinder stable alignment and consistent relationship learning across modalities, leading to inaccurate data recovery \rev{that fails to capture critical details.}

% generating the missing data 
% 将不同模态的分布transfer to a commonn space，然后根据模态不确定性恢复丢失数据，提高恢复数据的信息量。在同一空间对齐多模态特征，以消除模态分布变化对稳定的跨模态关系学习的影响
% we 将不同模态的分布transfer to a commonn space 来对齐多模态特征，从而提取稳定的跨模态关系
To address fluctuating modality discrepancies in data recovery, instead of directly learning cross-modality correlations based on varying feature distributions across modalities, we transfer the distributions of different modalities into \rev{a common space} to align modality features before consistent correlation learning as shown in Fig.~\ref{fig:training}.

% we reconstruct distribution of missing modality to restore the modality-specific feature representation with the correlations between modality distributions.
%% to \rev{learn the mapping relationship between their discriminative features which reflect to the modality characteristics according to the consistent semantic meaning.
%% we focus on the distribution transformation from the missing modality to the available modalities in feature space \rev{where} the missing data is recovered based on the transferred distribution of feature representations. 
% (简述过程)使用同一空间的模态特征来恢复丢失模态，生成的数据xxx, 因此移除了动态环境对模态特征对齐的影响
Normalizing modality distributions from available data to recover missing data bridges the gap between varying modality-specific features, facilitating the extraction of stable cross-modal correlations despite distribution fluctuations. 
% Such instability also hinders the recovery of missing data, as the mappings learned between modalities may no longer hold under fluctuating distributions. 
% By recovering data based on the cross-modality feature transformation from available data, the generated data contains the shared information among modalities and the complementary information of the missing modality as well, hence mitigating the distribution gap between the recovered and the true data.

% 为什么可以同一到同一空间，做法
% Despite the distinct features emphasized by each modality leading to inconsistent distributions, 
\nextrev{Different modalities share similar semantics as they capture complementary aspects of the same \rev{concept (e.g., a person’s health status)}.
This semantic similarity unleashes the potential for mapping their features to a common space where they are represented consistently.}
% distribution transformation between modalities by aligning their semantic representations. 
By transferring the distributions of different modalities into this common space, the model captures more stable and reliable correlations between the modalities without being affected by the fluctuations or discrepancies in their feature distributions.
% With the benefits of semantic consistency among modalities, \name transfers modality characteristics by learning the relationships between the distributions of different modalities.
% Specifically, 我们将所有现有模态的特征根据均值和方差，转换到正态分布，在此空间提取共同的语义信息
Specifically, we normalize the features $z_{i}^{m'}$ of all available modalities ($m'\ne m, m'\in \left\{ 1, ..., M \right\}$) with their means $\mu_{i}^{m'}$ and variances $\Sigma_{i}^{m',v}$ to a standard normal distribution $N(0, I)$, enabling the extraction of shared semantic information in the common space. After aligning modality for semantic information extraction, the cross-modality correlations are learned with a decoder $f_{\omega}^{m}(\cdot)$. Finally, we reconstruct the features of missing data $\hat{z}_{i}^{m}$ by leveraging both the shared semantic information and the complementary information from the available modalities as
\begin{equation}
% \hat{z}_{i}^{m} = \frac{1}{\sum_{m'} 1/{||\sigma^{m',v}_i||^2_2}} \sum_{m'} (\frac{z_{i}^{m'}-\mu_{i}^{m'}}{\sigma_{i}^{m',v}} \cdot \frac{1}{||\sigma^{m',v}_i||^2_2})
\hat{z}_{i}^{m} = f_{\omega}^{m} (u_{i}^{m}),~ u_{i}^{m} = \sum_{j \in m'} \left( \Sigma_i^{j,v} \right)^{-1/2}(z_{i}^{j}-\mu_{i}^{j})
\end{equation}

% 在获得语义信息后，我们训练一个decoder，根据共享的语义信息和丢失模态的互补信息，重建丢失数据。模态丢失重建的损失函数定义为MSE，可以表示为
To recover missing feature representations from the available modalities, the reconstruction loss is computed as \rev{the mean squared error  (MSE) between the recovered features and the original ones}:
\begin{equation} \label{eq:recover_loss}
\mathcal{L}_{recover} = \sum_{i=1}^{N} \left\| \hat{z}_{i}^{m} - {z}_{i}^{{m}} \right\|_{2}^{2}
\end{equation}

% 模态补偿后动态融合的过程
% 恢复的模态特征计算其对应的不确定性根据公式，然后和其他可用的模态特征一起输入动态权重分配模块来进一步提升多模态融合对关键状态识别的可靠性和敏感性。
% By learning inter-modal relationships in a common space, this approach mitigates cross-modal discrepancies arising from varying feature distributions, enhancing the integration of multimodal information for more reliable data recovery. 
The recovered modality features are first used to estimate its corresponding uncertainty with Eqn.~\eqref{eq:dynamic_fused_feature} and then fed into the transformer-based multimodal fusion module in Sec.~\ref{sec:transformer_fusion}, along with features from other available modalities, to further improve \needrev{reliability and sensitivity} in identifying critical health status. Finally, the overall training loss for outdoor health status detection is formulated as
\begin{equation}
\mathcal{L}_{total} = \mathcal{L}_{dyn} + \mathcal{L}_{recover}
\end{equation}

% % 为什么要做分布近似？为了显式表示数据的分布特征，以方便分布映射函数学习到特征的分布结构以重建模态 to minimize the distribution mapping error.
% To extract the complementary information of the missing modality for modality reconstruction, \name performs distribution approximation that captures the informative characteristics of feature distribution.
% % PCA可以提取少量重要特征来表征数据的分布，主成分反映出原始数据中最显著的变化方向，从而表示了数据的分布结构
% The well-known principal component analysis (PCA)~\cite{} indicates that a few critical principal components capture the most substantial directions of variance in the data, allowing for an appropriate estimation of the data's underlying structure without distinct discrepancy.
% Motivated by this, we perform PCA to the feature representations of the missing modality, and then capture the first $K$ important principal components (e.g., $K=4$) to keep the most significant information for distribution approximation.
% % 为什么要在特征域进行分布近似？捕获了数据的关键特征和模式，在特征域进行分布估计可以更好地理解数据的特征和结构，识别和选择最具信息量的特征。经过feature encoder的特征缩放、特征归一化等，这些预处理步骤可以提高模型的稳定性和收敛速度
% % Since the data after feature extraction facilitates a better understanding of key characteristics for data distribution, we estimate the distribution in the feature space to identify the most informative structure.

% \begin{figure}[t] 
% \centering
% \includegraphics[width=\linewidth]{figures/Modality_Reconstruction.pdf}
% \caption{Modality reconstruction module with normalized modality distributions.}
% \label{fig:distribution_reconstruction}
% \vspace{-2ex}
% \end{figure}

\vspace{-0.1cm}
% 0.5 page
\section{Experimental Setup} \label{sec:implementation}
In this section, we demonstrate the detailed experimental setup of our \name system for outdoor health monitoring using {Stressors} dataset~\cite{taamneh2017multimodal} and UP-Fall Detection dataset~\cite{martinez2019up}. The performance of \name is evaluated against several multimodal fusion algorithms using carefully selected hyper-parameters to ensure a fair comparison.

{\textit{1) \textbf{Dataset and tasks.}}}
% task (facial+vital)
% data type(device), location, resolution
% pre-processing(meaning and shape), label/ground truth
% samples - train dataset, test dataset
\rev{Public datasets explicitly designed for outdoor health monitoring with physiological and cardiac indicators are extremely limited.}
Therefore, we adopt two representative health-related multimodal datasets to evaluate the performance of \name, the {Stressors} dataset~\cite{taamneh2017multimodal} for \needrev{stress recognition} and the UP-Fall Detection dataset~\cite{martinez2019up} for human fall detection. 
% The Stressors dataset analyzes drivers' status under various stressors, enabling the identification of stress levels experienced in a vehicle. 
The Stressors dataset captures dynamic physiological and emotion changes experienced by drivers under real-world stress-inducing conditions, such as dense traffic and secondary distractions.
Since stress-induced physiological responses such as elevated heart rate and irregular breathing are early indicators of health risk~\cite{nvemcova2020multimodal}, 
% especially during cognitively demanding tasks like driving.
% \needrev{Given the similarity between stress patterns and early signs of abnormal health conditions~\cite{nvemcova2020multimodal}}, 
we utilize facial information, physiological signals (e.g., heart rate and breathing rate), and vehicle parameters (e.g., speed, acceleration, brake force, steering angle, and lane position) to detect stress-related abnormalities for the driver.
Facial data is captured at 25 fps, while physiological signals and vehicle parameters are sampled at 1 Hz. The multimodal data is synchronized using global timestamps and segmented into 10-second windows with 5-second overlap. 
% Due to the lack of ground truth, labels are manually annotated based on facial video and palm EDA sensor data.
Finally, a total of 1500 samples from 24 subjects are selected with 4 subjects' data for testing and the others for training.

% \newrev{Recognizing activities is more complex than status recognition, as it requires understanding temporal sequences and spatial movements. Therefore, we further evaluate the performance of our \name on activity recognition.}
% % The Drive\&Act dataset is a multimodal driving behavior dataset for detecting 34 fine-grained distractive activities, such as fetching objects, looking around, and talking on the phone. Our experiments utilize RGB video from the right-top view and 3D skeleton data, where RGB video is sampled at 30 fps and 3D pose at 25 Hz. 
% % Data streams are also segmented into 10-second windows with a 5-second overlap, with 75\% and 25\% of 15 subjects' data allocated for training and testing.
% The UTD dataset~\cite{chen2015utd} is a multimodal human action dataset from multiple sensors (depth cameras, accelerometers, and gyroscopes), which includes 27 different human actions, covering common daily activities such as waving, drinking, punching, and kicking. 
% Our experiments utilize the 3D skeleton data and inertial data, where the sampling rate is 30 Hz and 50 Hz, respectively. The skeleton contains 20 joints and the inertial data includes acceleration and rotation signals. We use 6 subjects' data for training and 2 subjects' data for testing.

\rev{We also utilize the UP-Fall Detection dataset~\cite{martinez2019up}, which focuses on detecting health crises associated with abrupt physical incidents such as falls.
Fall events are critical health emergencies, particularly for elderly populations, and are often preceded by abnormal behavioral or physiological patterns.}
% UP-Fall Detection Dataset~\cite{martinez2019up} is a publicly available multimodal dataset designed for human fall detection and activity recognition systems, which 
The dataset contains recordings from 17 participants performing 11 daily activities, including 5 types of falls (e.g., falling forward using hands, falling sideward, and falling sitting in empty chair) and 6 non-fall activities. It incorporates data from multiple synchronized sensors, including three-axis accelerometers and gyroscopes sampled at 100 Hz, placed on the waist, wrist, and left ankle, along with RGB video at 18 Hz.
% Although the dataset lacks physiological signals such as heart rate or respiration, it offers valuable information for modeling sudden, high-risk health events that may follow behavioral instability.
\newrev{We follow the \rev{subject-independent protocol} by using data from 12 participants for training and other 4 participants for testing whose physiological and behavioral patterns are differ significantly.}

{\textit{2) \textbf{Baselines.}}}
% baselines and evaluation metrics (MAE/GAN, supervised multimodal fusion for driver, Contrastive Multi-view Learning-CMC)
To investigate the advantages of our \name framework, we compare it with the following multimodal fusion benchmarks:
\begin{itemize}
    \item {\bf{DeepSense~\cite{yao2017deepsense}}} is a unified deep learning framework for general multimodal sensing applications like driver monitoring, which integrates convolutional neural networks for extracting spatial features and recurrent neural networks for capturing temporal dependencies. 
    \item {\bf{MAP~\cite{ji2023map}}} is a novel vision-language pre-training framework that incorporates uncertainty quantification into multimodal semantic understanding. The framework dynamically adjusts multimodal fusion based on inter-modal uncertainty derived from the probabilistic distributions of each modality's representations.
    \item {\bf{Missing Modality Imagination Network (MMIN)~\cite{zhao2021missing}}} is a unified model for multimodal emotion recognition in scenarios with uncertain missing modality, where two independent networks are employed to reconstruct the missing modality based on other available modalities in the forward direction and also predict the available modalities based on the imagined missing modality in the backward direction.
    \item {\bf{Health-LLM~\cite{kim2024health}}} is a specialized medical-domain LLM framework designed to address the challenges posed by high-dimensional, non-linear, and non-linguistic time-series data in the healthcare domain. The integration of health-specific knowledge into prompts enables effective interpretation of complex patterns in multimodal data like physiological and behavioral signals.
    % bridge the gap between general pre-trained LLMs and the specific health prediction tasks 
\end{itemize}

{\textit{3) \textbf{Models and hyper-parameters.}}}
We train our \name on a server with an NVIDIA RTX 5000 GPU of 32\!~GB, Intel i9-10885H CPUs, and 256\!~GB RAM. 
% Python 3.8 and PyTorch 1.13.1 are used for implementing the application.
% The facial information, physiological signals, and vehicle data are processed to dimensions of (60, 50, 8), (120, 8), and (120, 6) for training, respectively, with the extracted feature dimension of 1280.
\waitrev{For MAP~\cite{ji2023map}, the ViLT model~\cite{kim2021vilt} is used as the backbone with BERT model for language prompts.
For Health-LLM~\cite{kim2024health} and our \name, we adopt the MedAlpaca-7B model as the pre-trained backbone and perform instruction fine-tuning using 8-rank LoRA on both datasets. %The ViT-B/32 model is used as the backbone to extract visual features.
Time-series data are converted into textual prompts following the format used in Health-LLM, incorporating heartbeat, breathing rate, and facial emotion for the Stressors dataset, and accelerometer and gyroscope data for the UP-Fall Detection dataset.}
We employ the same LSTM-based feature encoders for MMIN~\cite{zhao2021missing} and similar CNN network for multimodal feature learning in DeepSense~\cite{yao2017deepsense}.
% 没说模态的特征提取用的什么模型
% We follow the original experimental settings of MAP~\cite{ji2023map} for fine-tuning. Similarly, we adopt the same LSTM based feature encoders with a modality-complete base model for MMIN as described in~\cite{zhao2021missing}. 
% For Health-LLM~\cite{kim2024health} and our \name, We set the optimizer, learning rate and batch size as Adam, 0.0001 and 2, respectively. 
% Due to the limited 32\!~GB memory of the NVIDIA RTX 5000, \rev{we fine-tune the large language model using 8-rank LoRA with 4-bit data quantization~\cite{dettmers2024qlora}} and a maximum sequence length of 960 to ensure complete input information.
Moreover, we use a 4 transformer layers for task head and 2 layers for \rev{learning mean and variance of feature probabilistic distribution}.
The number of layers in transformer-based multimodal fusion is set to 4 for cross-modal correlation extraction.
\rev{For the modality reconstruction network, we employ a traditional autoencoder with residual connections, where the encoder consists of 4 transformer layers and the decoder comprises 4 transposed convolutional layers.}
The hyper-parameters for regularization of unimodal modal and uncertainty calibration are set to $\lambda_u=0.1$ and $\lambda_c=0.1$, respectively.
% \needrev{The number of important principal components is set to $K=4$ by default unless specified otherwise.}
% To demonstrate the training effectiveness of different methods, we run each experiment five times and report the model performance on the testing set during training.
\waitrev{\newrev{Following prior work~\cite{chang2020data, gao2024embracing, maddox2019simple}, we inject various kinds of noises (e.g., background noise, lighting variations, and occlusions) into both datasets to simulate realistic outdoor environments.} By default, 50\% of multimodal samples are injected with noise, while the rest remain high-quality.}
% \needrev{To simulate the realistic outdoor environments, we randomly inject noise into 50\% of the samples from a single modality while leaving data from other modalities in high quality, and then we assume 50\% facial data missing by default.} 动态环境变化没有体现
% We run each experiment five times and report the average performance on the testing set.

% In the experiments, we deploy $N$ satellites orbiting the Earth, distributed across $J$ different altitude orbits. $N$ and $J$ are set to 24 and 3 by default unless specified otherwise. The default orbital period ratio is  $p_1:p_2,...,:p_J=1:1.1:1.3$. The computing and memory resource budgets among satellites follow a discrete uniform distribution (i.e., $P(b_i=0.25, 0.5, 0.75, 1)=0.25$). The uplink rates are set based on real-world traces~(e.g., RTT) collected from Starlink.  

% 3 pages (Figure+Table 8~12)
% setting, (observation, reason/conclusion)*2
\section{Evaluation} \label{sec:evaluation}
In this section, we evaluate the overall performance of our \name framework and various benchmarks. We also evaluate the performance of the proposed framework under different levels of data quality degradation and dynamic modality adaptation.
The contributions of different modules within our \name framework are also analyzed to illustrate their individual roles in the proposed framework.
% \rev{This section provides numerical results to evaluate the training performance of \name framework and the effectiveness of each meticulously designed component.}

\begin{figure}[t]
\vspace{-.5ex}
\centering
\subfloat[Poor lighting or occlusion \label{fig:accuracy_label_stressor}]{
    \includegraphics[width=0.49\linewidth]{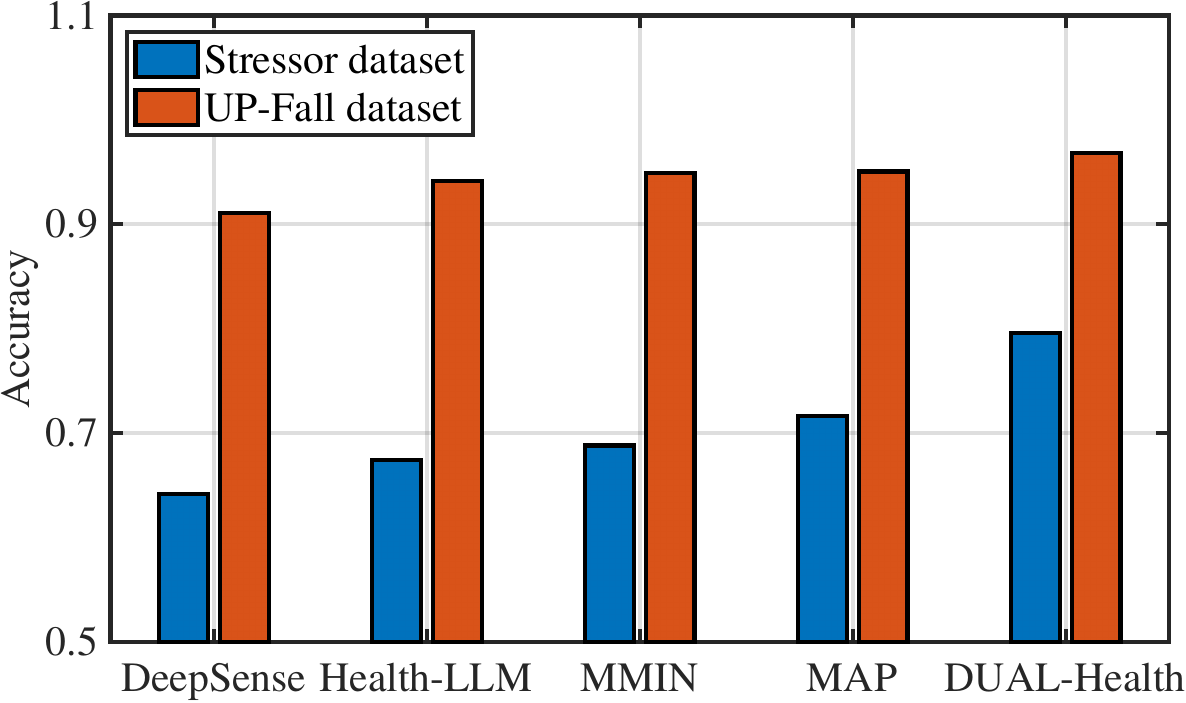}
}
% \hspace{.1cm}
\subfloat[High background noise \label{fig:accuracy_label_UTD}]{
    \includegraphics[width=0.49\linewidth]{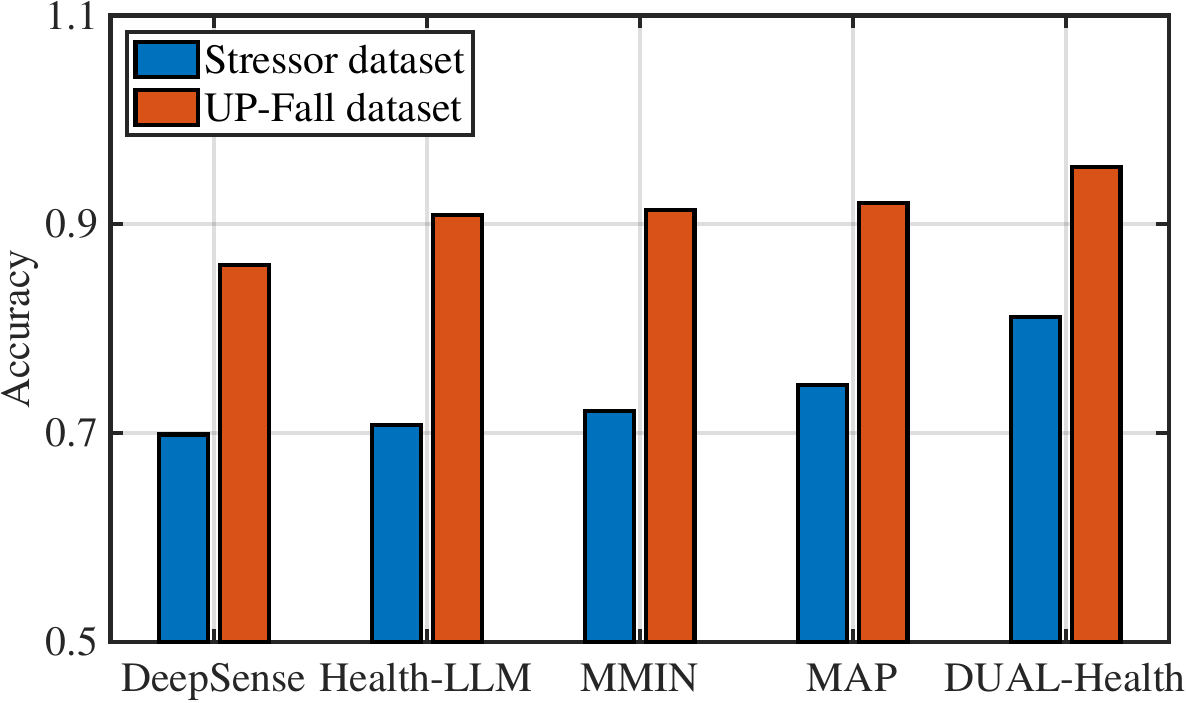}
}
\caption{Accuracy of baselines with 50\% noisy inputs under outdoor conditions involving \needrev{background noise}, lighting variations, and occlusions.}
\label{fig:accuracy_label}
\vspace{-1.5ex}
\end{figure}

\begin{figure*}[htbp!]
\setlength\abovecaptionskip{6pt}
\subfloat[Precision]{
    \centering
    \label{fig:metrics_pre}
    \includegraphics[width=0.223\linewidth, height=2.36cm]{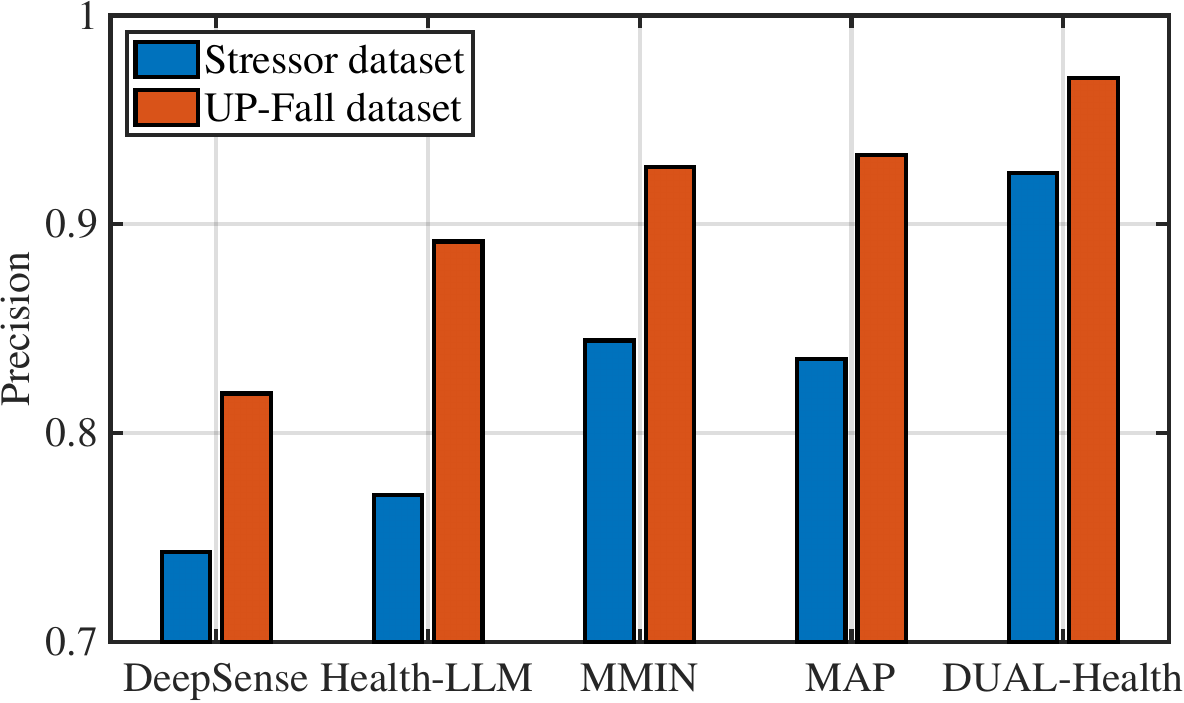}
}
\subfloat[Recall]{
    \centering
    \label{fig:metrics_rec}
    \includegraphics[width=0.223\linewidth, height=2.36cm]{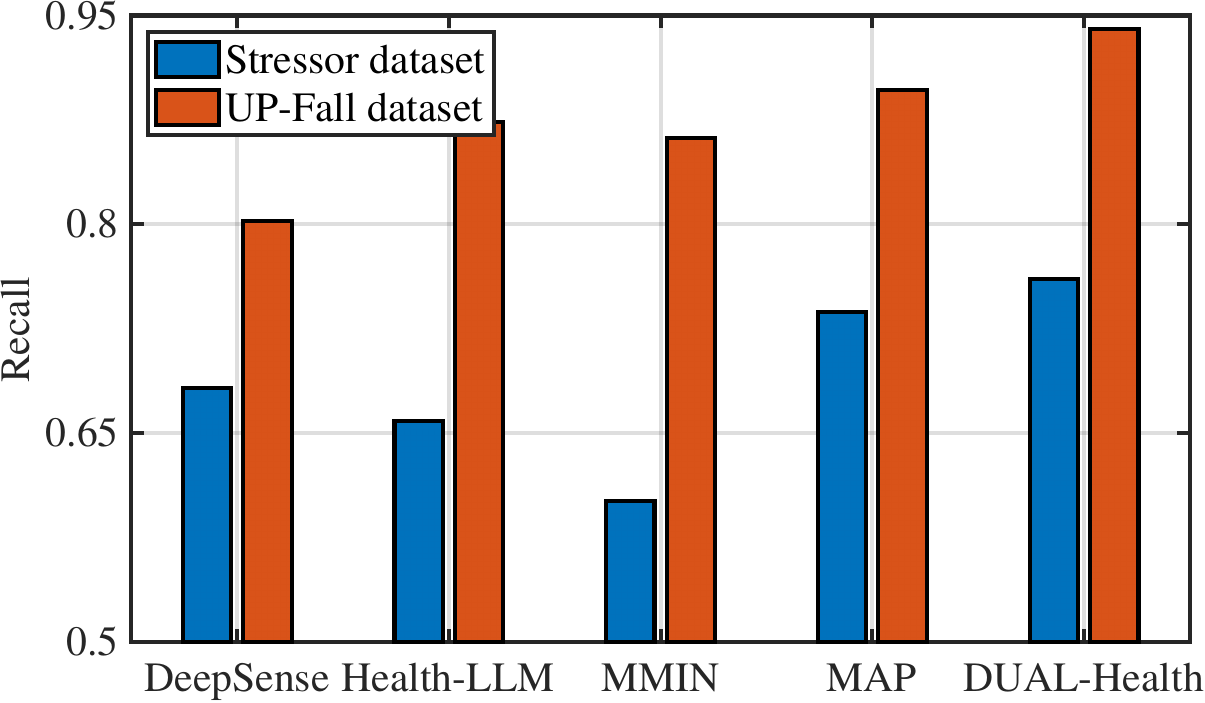}
}
\subfloat[F1 score]{
    \centering
    \label{fig:metrics_f1}
    \includegraphics[width=0.223\linewidth, height=2.36cm]{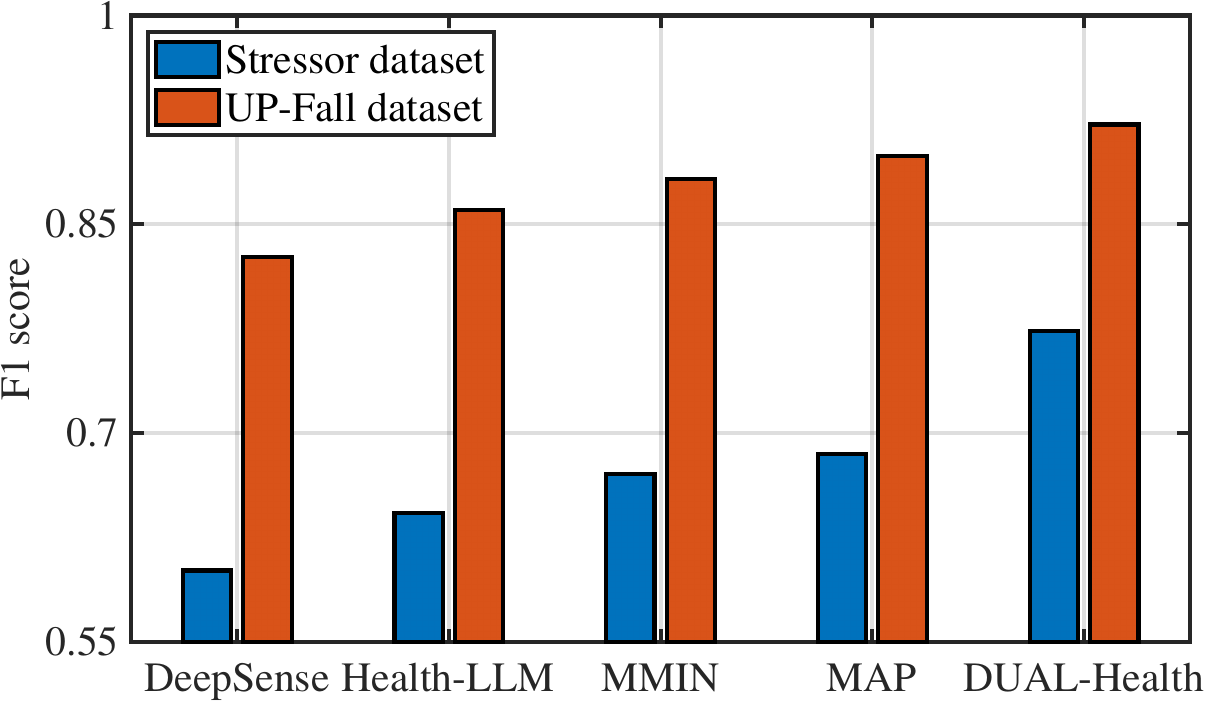}
}
\subfloat[AUROC]{
    \centering
    \label{fig:metrics_auroc}
    \includegraphics[width=0.223\linewidth, height=2.36cm]{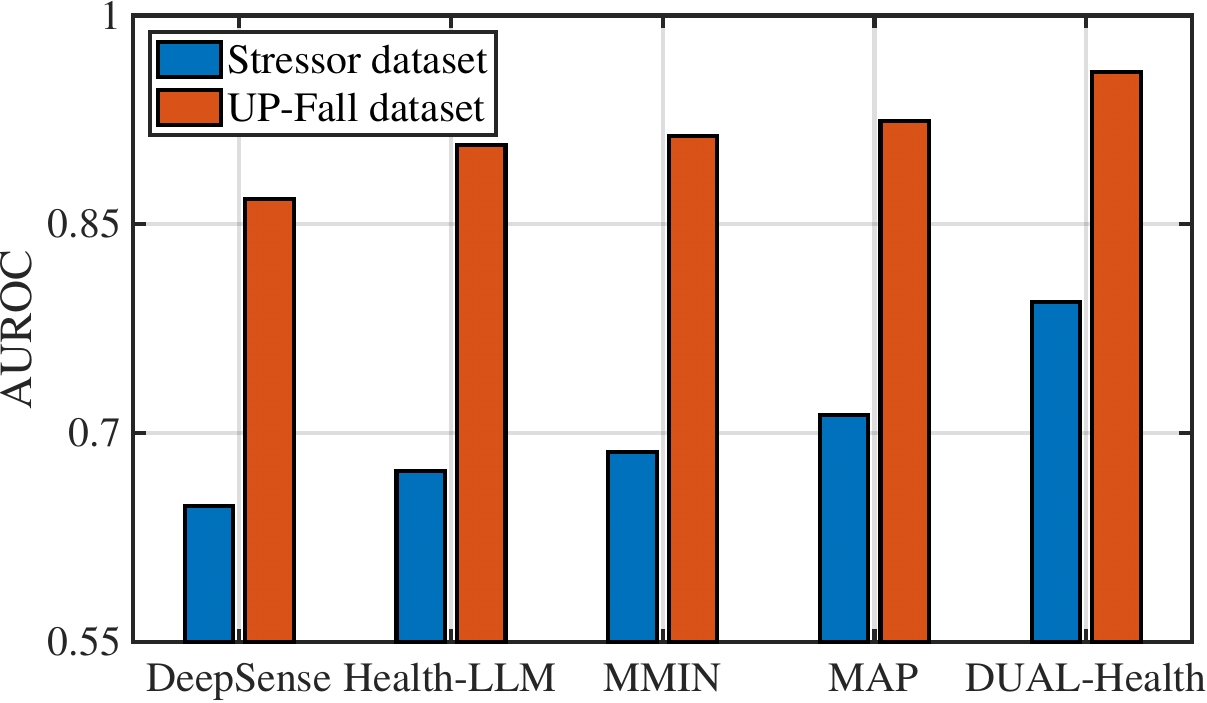}
} \\
\vspace{-1ex}
\subfloat[Precision]{
    \centering
    \label{fig:metrics_pre}
    \includegraphics[width=0.223\linewidth, height=2.36cm]{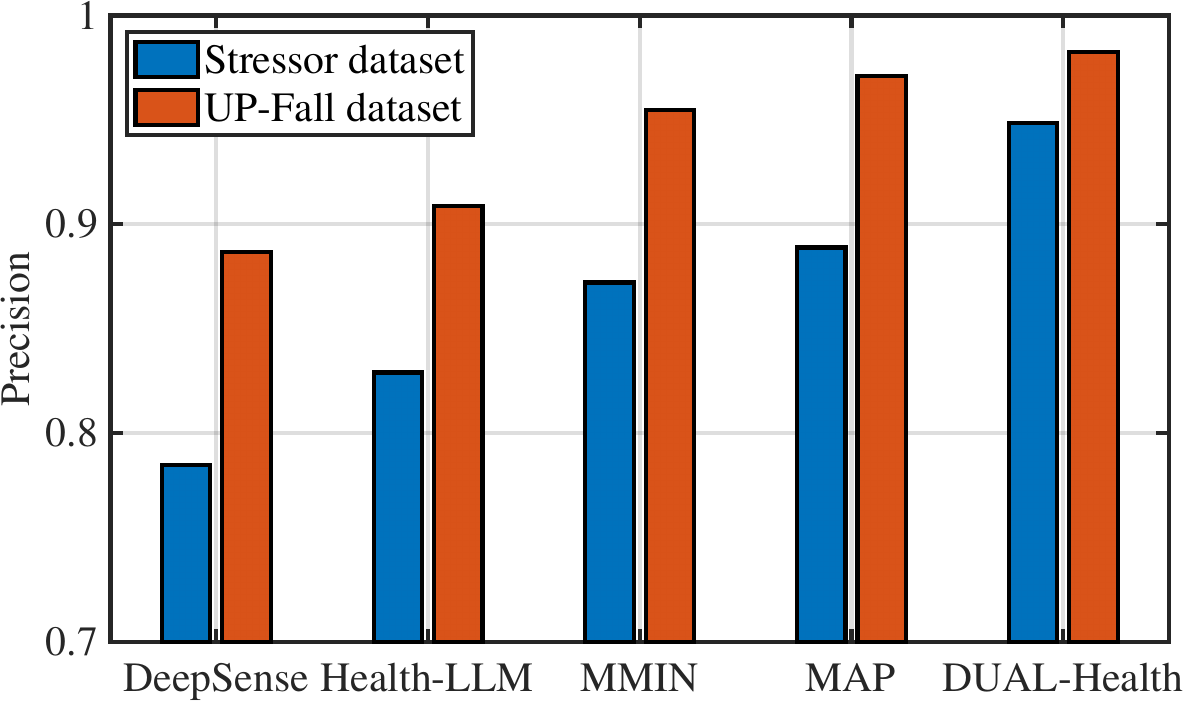}
}
\subfloat[Recall]{
    \centering
    \label{fig:metrics_rec}
    \includegraphics[width=0.223\linewidth, height=2.36cm]{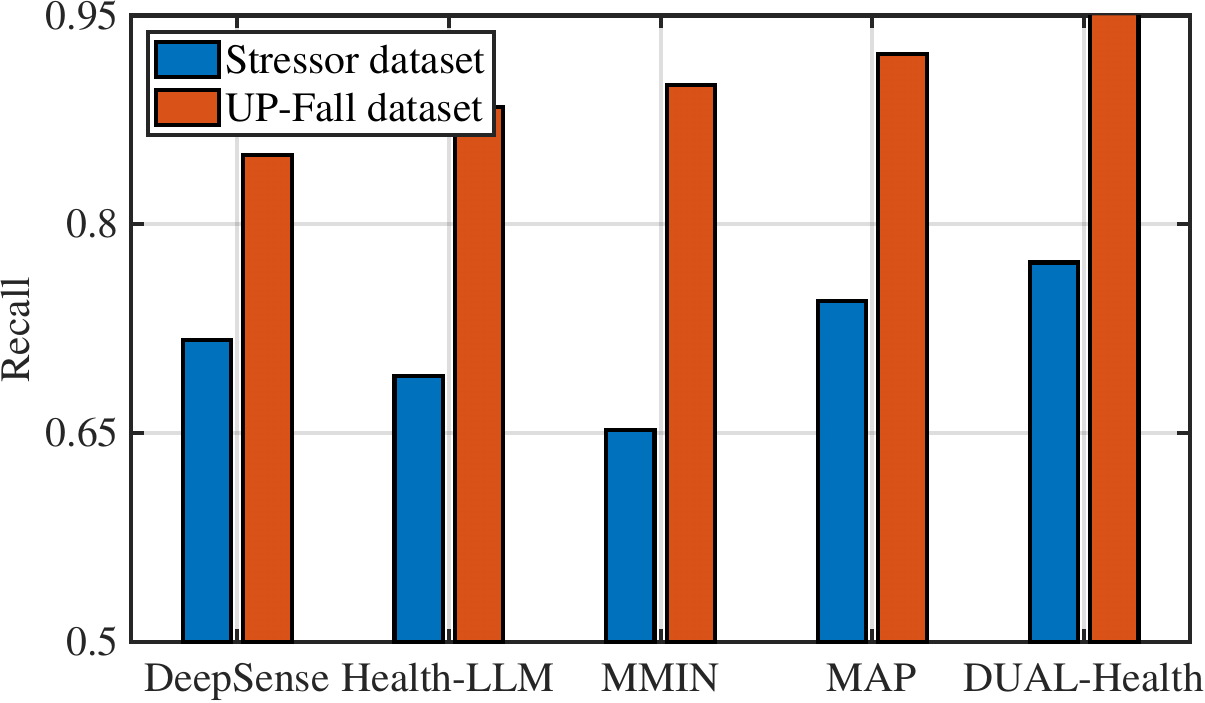}
}
\subfloat[F1 score]{
    \centering
    \label{fig:metrics_f1}
    \includegraphics[width=0.223\linewidth, height=2.36cm]{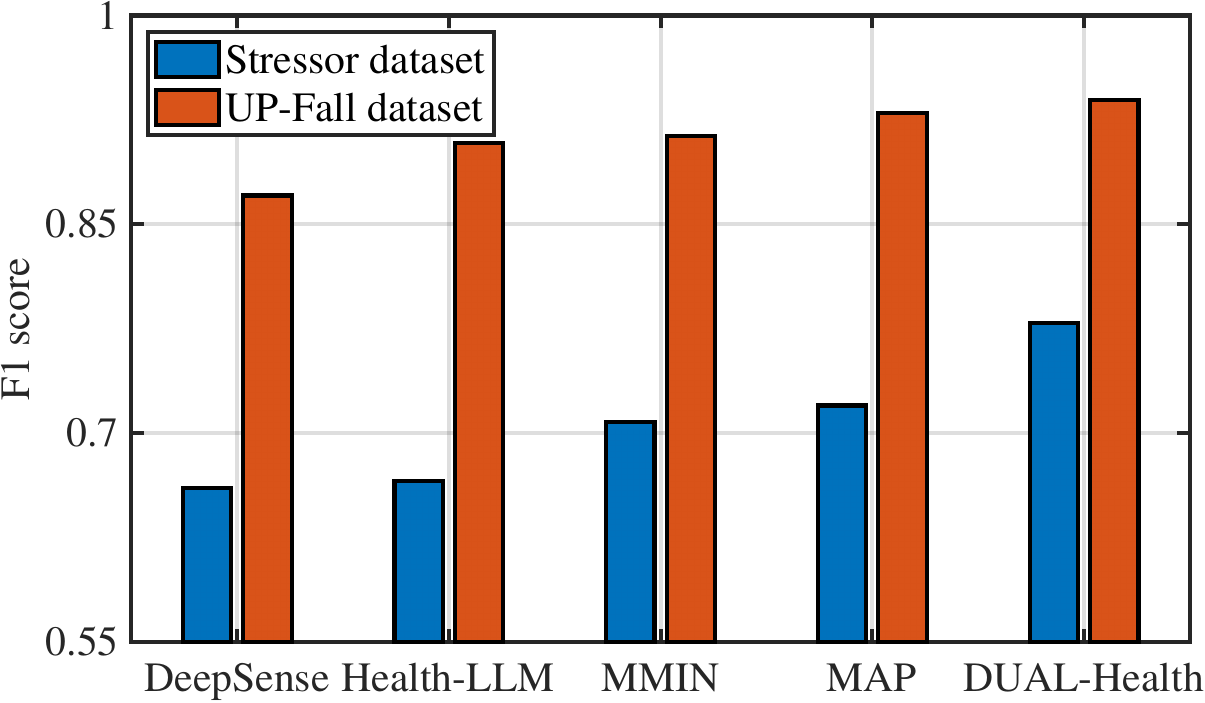}
}
\subfloat[AUROC]{
    \centering
    \label{fig:metrics_auroc}
    \includegraphics[width=0.223\linewidth, height=2.36cm]{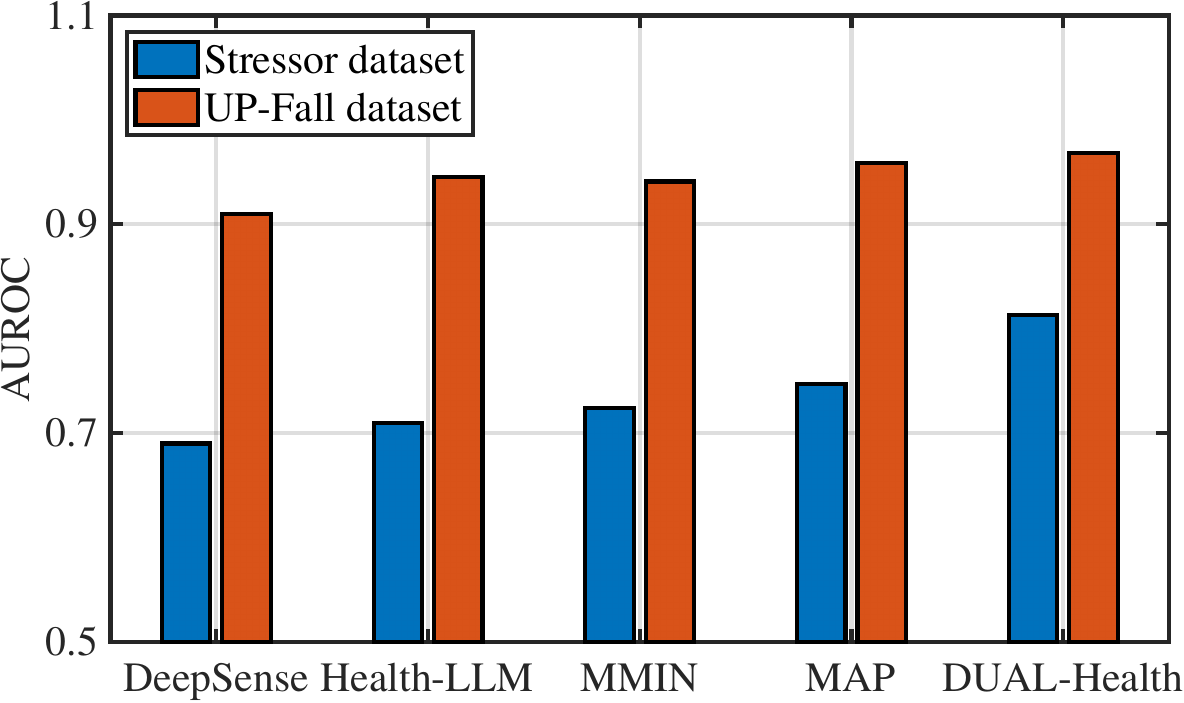}
}
% \vspace{-.5ex}
\caption{Class-specific metrics for baselines with 50\% low-quality inputs under poor lighting or occlusions (Fig. (a)-(d)) and high background noise (Fig. (e)-(h)).}
\label{tab:accuracy_metrics}
\end{figure*}

% % & 77.50 & 91.42 & 74.54 & 74.26 & 75.94 & 90.46 & 76.91 & 72.85 & 74.65 & 87.70 & 77.15 & 71.99 \\
\begin{table*}[t]
\centering
\scalebox{0.95}{
\begin{tabular}{c|c|cccc|cccc|cccc}
    \hline 
    % Model & Acc & Pre & Rec & F1 & Acc & Pre & Rec & F1 & Acc & Pre & Rec & F1 \\
    \multirow{2}{*}{Missing} & \multirow{2}{*}{Model} & \multicolumn{4}{c|}{0\% noisy}  & \multicolumn{4}{c|}{50\% noisy} & \multicolumn{4}{c}{100\% noisy} \\
     & & {Acc} & {Pre} & {Rec} & {F1} & {Acc} & {Pre} & {Rec} & {F1} & {Acc} & {Pre} & {Rec} & {F1} \\
    \hline 
    \multirow{5}{*}{0\%} 
    & {DeepSense} & 71.04 & 79.08 & 72.39 & 65.21 & 66.87 & 78.32 & 68.54 & 63.53 & 64.30 & 75.08 & 67.16 & 60.57 \\
    & {Health-LLM} & 73.51 & 84.38 & 73.04 & 70.01 & 71.84 & 81.17 & 70.11 & 66.56 & 69.04 & 77.56 & 67.78 & 62.89 \\
    & {MMIN} & 72.33 & 90.38 & 62.56 & 71.88 & 70.82 & 88.23 & 60.57 & 69.74 & 69.46 & 87.23 & 61.21 & 66.84 \\
    & {MAP} & 77.05 & 86.34 & 74.12 & 73.22 & 74.42 & 84.25 & 72.66 & 70.43 & 71.94 & 82.55 & 69.48 & 67.97 \\
    & {\name} & \textbf{82.31} & \textbf{95.88} & \textbf{78.22} & \textbf{78.16} & \textbf{81.75} & \textbf{94.72} & \textbf{77.26} & \textbf{77.44} & \textbf{80.98} & \textbf{94.43} & \textbf{76.92} & \textbf{77.03} \\
    \hline 
    \multirow{5}{*}{50\%} 
    & {DeepSense} & 67.44 & 78.59 & 69.21 & 63.74 & \needrev{64.17} & 74.32 & 68.27 & \needrev{60.14} & 61.24 & 70.99 & 60.25 & 57.35 \\
    & {Health-LLM} & 70.01 & 80.24 & 68.53 & 66.37 & 67.39 & 77.06 & 65.89 & 64.24 & 64.59 & 73.15 & 63.71 & 61.46 \\
    & {MMIN} & 70.58 & 87.31 & 62.14 & 69.22 & 68.82 & 84.44 & 60.14 & 67.06 & 67.16 & 83.06 & 56.31 & 64.64 \\
    & {MAP} & 74.06 & 85.27 & 74.04 & 70.94 & 71.63 & 83.56 & 73.70 & 68.53 & 68.12 & 79.47 & 68.06 & 64.59 \\
    & {\name} & \textbf{80.66} & \textbf{93.74} & \textbf{76.67} & \textbf{76.83} & \textbf{79.89} & \textbf{94.47} & \textbf{76.08} & \textbf{77.35} & \textbf{79.14} & \textbf{92.81} & \textbf{75.26} & \textbf{75.74} \\
    \hline 
  \end{tabular}
}
\caption{Performance comparison for baselines under different levels of data quality degradation with/without 50\% facial data missing on the Stressors dataset.}
\label{tab:noise_rate}
\vspace{-1ex}
\end{table*}

\vspace{-0.1cm}
\subsection{The Overall Performance}
{\textit{1) \textbf{Detection Accuracy.}}}
Fig.~\ref{fig:accuracy_label} demonstrates the detection accuracy of \name and other baselines in health monitoring under 50\% noisy inputs and 50\% facial information missing on the {Stressors} and UP-Fall Detection datasets. 
Our \name framework outperforms all other baselines under various outdoor environments, primarily attributing to its precise estimation of each modality's contribution from noisy or incomplete inputs, thus enabling reliable and timely health monitoring through the proposed transformer-based multimodal fusion.
By reconstructing missing modality which is resilient to modality distribution fluctuations and estimating the dynamic uncertainty of each input, \name adaptively relies on more reliable and sensitive input data. This results in an accuracy improvement of \needrev{15\% and 11\%} over DeepSense and MMIN, respectively.
Although Health-LLM and MAP also employ MLLMs to harness general medical knowledge and identify subtle health changes, \name exhibits superior detection accuracy. This is because it balances the reliability of varying data quality with the fluctuations of health biomarkers, ensuring that subtle health changes can be detected without being overshadowed by low-quality data. Furthermore, by prioritizing high-contributing modalities in transformer-based feature fusion, \name strengthens model ability to focus on critical cross-modal correlations, further enhancing detection accuracy.
% leading to more accurate detection of abnormal health status.

% \begin{table*}[htbp]
% \centering
% % \scalebox{0.9}{
%   \begin{tabular}{c|c|cccccc}
%     \hline
%     Dataset & Model & Accuracy & Precision & Recall & F1 score & Loss & AUROC \\
%     \hline
%     \multirow{5}{*}{Stressors} 
%     & {DeepSense} & \needrev{64.17} & 74.32 & 68.27 & \needrev{60.14} & 0.820 & \needrev{65.92} \\
%     & {Health-LLM} & 67.39 & 77.06 & 65.89 & 64.24 & 0.661 & 68.28 \\
%     & {MMIN} & 69.12 & 87.44 & 60.14 & 70.06 & 0.527 & 70.48 \\
%     & {MAP} & 71.63 & 83.56 & 73.70 & 68.53 & 0.345 & 72.59 \\
%     & \textbf{{\name}} & \textbf{79.89} & \textbf{94.47} & \textbf{76.08} & \textbf{77.35} & \textbf{0.108} & \textbf{80.64} \\
%     \hline
%     \multirow{5}{*}{UTD} 
%     & {DeepSense} & \needrev{69.28} & 78.62 & 73.91 & \needrev{64.13} & 0.722 & \needrev{70.66} \\
%     & {Health-LLM} & 72.78 & 83.56 & 72.88 & 68.91 & 0.386 & 73.69 \\
%     & {MMIN} & 73.12 & 93.73 & 65.14 & 72.64 & 0.402 & 74.38 \\
%     & {MAP} & 75.34 & 87.56 & 73.78 & 71.70 & 0.252 & 76.32 \\
%     & \textbf{{\name}} & \textbf{85.12} & \textbf{94.27} & \textbf{74.84} & \textbf{77.13} & \textbf{0.102} & \textbf{86.56} \\
%     \hline
%   \end{tabular}
%   % }
%   \caption{Class-specific metrics for baselines with 50\% low-quality inputs and 50\% facial information missing} \label{tab:accuracy_metrics}
%   \vspace{-1ex}
% \end{table*}

{\textit{2) \textbf{Class-specific Metrics.}}}
To provide a more comprehensive evaluation of model performances in detecting abnormal health status, we compare class-specific metrics in Fig.~\ref{tab:accuracy_metrics}, including accuracy, precision, recall, F1 score, and AUROC (Area Under the Receiver Operating Characteristic Curve).
As shown in Fig.~\ref{tab:accuracy_metrics}, \name surpasses DeepSense, Health-LLM, MMIN, and MAP in precision by nearly \needrev{17\%, 13\%, 8\%, and 8\%}, respectively. This improvement stems from the proposed adaptive modality weight assignment module, which jointly accounts for input and fluctuation uncertainty in dynamically changing environments, thereby minimizing the interference of low-quality modalities to multimodal fusion while making the detections trustworthy.
We also notice that the proposed framework achieves the highest recall, approximately \needrev{77\% on the Stressors dataset and 94\% on the UP-Fall Detection dataset}, highlighting its ability to \waitrev{recover missing data across low-quality modalities and calibrate each input’s uncertainty in line with its relative contribution to the detection of health status.} 
This enables \name to promptly capture subtle yet critical health changes in health biomarkers under varying data quality and biomarker fluctuations, thereby facilitating the early detection of potential health issues.
Other MLLM benchmarks, in contrast, lacking customized design for uncertainty quantification and modality reconstruction to handle low-quality data, prone to over-relying on irrelevant information and overlooking critical health issues in dynamic environments.
Moreover, it is noteworthy to observe that our \name has a significantly higher F1 score and AUROC, further underscoring its sensitivity and reliability in timely identifying abnormal health status. 

\vspace{-0.1cm}
\subsection{Micro-benchmarking}
{\textit{1) \textbf{The Impact of Varying Data Quality.}}}
Table~\ref{tab:noise_rate} investigates the performance of health detection for our \name and other benchmarks \needrev{under different levels of data quality degradation with/without 50\% facial data missing on the Stressors dataset.} 
We notice that \name consistently exhibits the best detection performance across varying data quality in both scenarios compared to other benchmarks. 
This is attributed to its dynamic adjustment of modality-specific weights in multimodal fusion in the transformer layers and its stable multimodal alignment for missing data recovery with modality distribution fluctuations.  
The proposed framework calculates the dynamic contributions of low-quality data in changing environments, effectively leveraging multimodal complementary information to extract critical discriminative characteristics to mitigate performance degradation.  
Consequently, in the scenarios without data missing, \name maintains a stable accuracy exceeding \needrev{80\%} even with 100\% low-quality inputs, experiencing only a slight drop of \needrev{1.33\%} compared to that under no-degradation condition.  
In contrast, DeepSense, MMIN, and Health-LLM struggle to learn informative features for classification due to the lack of adaptive modality-specific weight assignment. Their accuracy drops sharply to lower \needrev{70\%}, and F1-score declines by over \needrev{5\%} \rev{when 100\% of the inputs contain noise}. MAP, on the other hand, fails to differentiate indicator fluctuations in varying-quality inputs, leading to inaccurate uncertainty estimation and weakened recognition of health indicator changes.

Moreover, the performance gap between \name and the other benchmarks is much larger in the scenario with 50\% missing data. \name achieves nearly \needrev{81\%} accuracy and \needrev{77\%} F1 score under no data degradation and surpasses DeepSense, Health-LLM, MMIN, and MAP by \needrev{18\%, 15\%, 12\%, and 11\%} accuracy on 100\% low-quality inputs, respectively. With the alignment of low-quality modalities through the common semantic space, \name learns consistent cross-modal correlations to reconstruct missing modalities.  
However, neglecting modality distributions normalization in other benchmarks results in notably inferior performance under no modality missing compared to the scenario under 50\% facial data missing. Although MMIN is capable of recovering missing data from other modalities, overlooking the adverse impact of noise on cross-modal alignment limits its performance, which highlights the superior adaptability of \name to \rev{dynamic environments with low-quality data}.

\begin{figure}[t]
% \vspace{-.5ex}
\setlength\abovecaptionskip{6pt}
% \centering
\subfloat[Input and fluctuation noise]{
    \centering
    \label{fig:dynamic_weight_noise}
    \includegraphics[width=0.48\columnwidth, height=2.36cm]{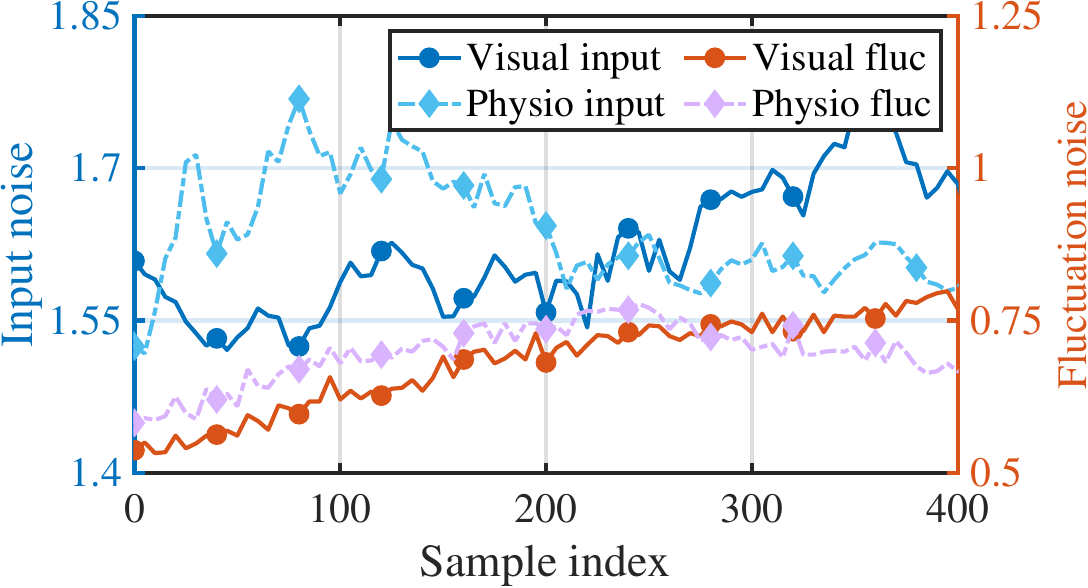}
} \hspace{-1ex}
\subfloat[Input and fluctuation noise]{
    \centering
    \label{fig:dynamic_weight_noise1}
    \includegraphics[width=0.472\columnwidth, height=2.36cm]{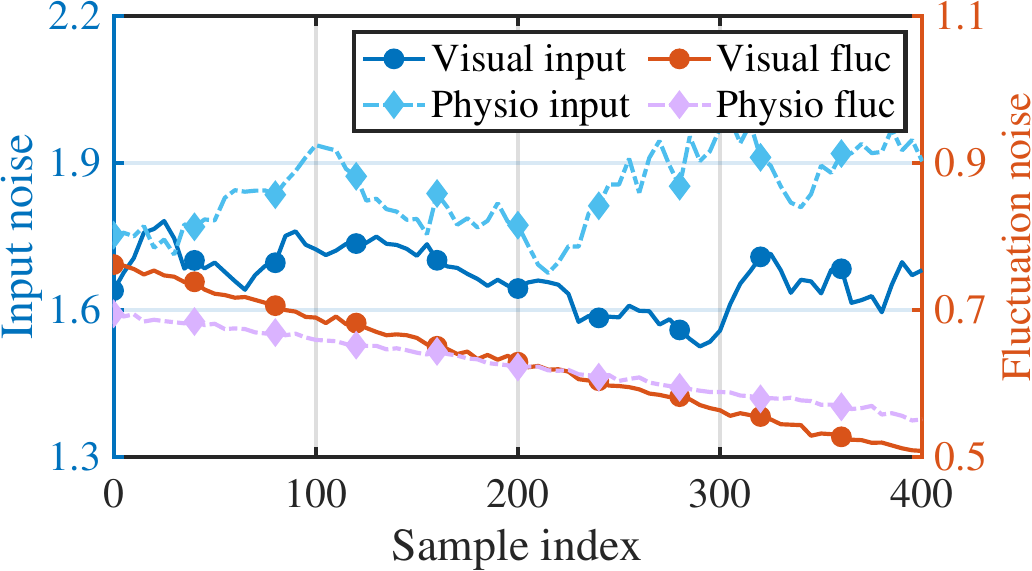}
} \\
\subfloat[Fusion weights with poor lighting in first 200 samples and high background noise in last 150 samples.]{
    \centering
    \label{fig:dynamic_weight_modality}
    \includegraphics[width=0.445\columnwidth, height=2.37cm]{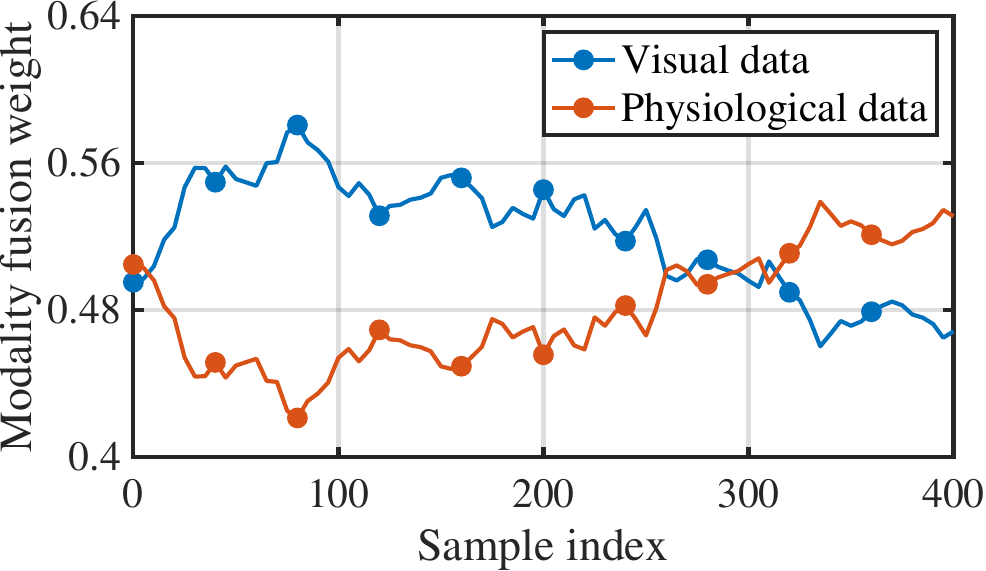}
} \hspace{.5ex}
\subfloat[Fusion weights with abrupt heartbeat changes in sample 80 to 120 and background noise in last 200 samples.]{
    \centering
    \label{fig:dynamic_weight_modality1}
    \includegraphics[width=0.447\columnwidth, height=2.36cm]{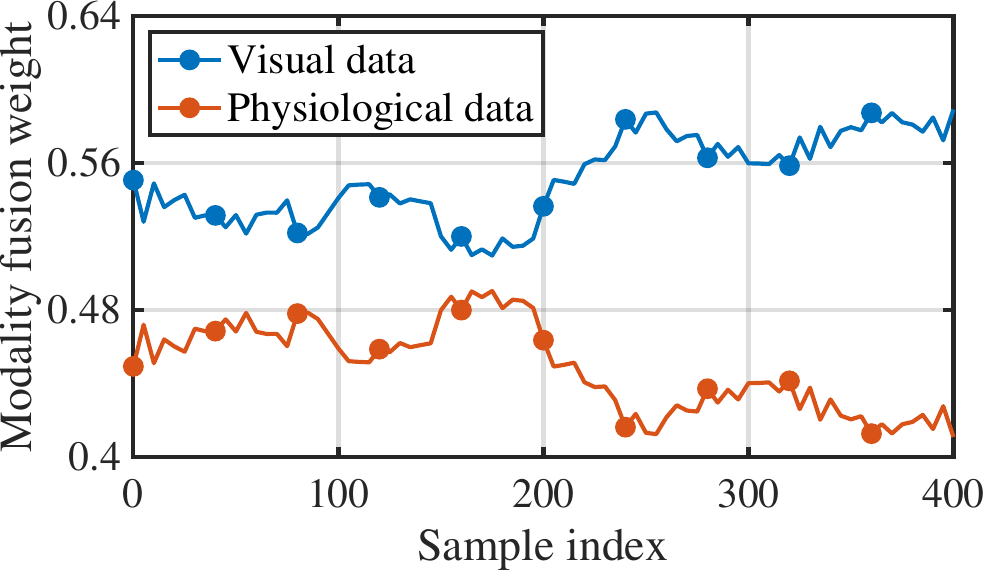}
}
\caption{\waitrev{Dynamic adaptation of \name on the Stressor dataset with changing environments.}}
\label{fig:dynamic_weight}
\vspace{1ex}
\end{figure}

{\textit{2) \textbf{Dynamic Adaptation of Fusion Weights.}}}
\waitrev{Fig.~\ref{fig:dynamic_weight} illustrates the estimated noise value and modality-specific fusion weights of \name under dynamically changing lighting and background noise conditions on the Stressors dataset.
As shown in Fig.~\ref{fig:dynamic_weight_noise} and Fig.~\ref{fig:dynamic_weight_modality}, modality-specific fusion weights remain low when the corresponding input quality deteriorates, thereby preventing low-quality modalities from adversely impacting overall model performance. This adaptive adjustment highlights \name's ability to dynamically prioritize reliable inputs, enabling adaptive and robust multimodal fusion under changing data quality.
% Figure~\ref{fig:dynamic_weight_facial_acc} and Figure~\ref{fig:dynamic_weight_phy_acc} further demonstrates that the accuracy of unimodal classification is closely aligned with modality-wise fusion weights, where higher unimodal accuracy corresponds to greater modality weights. 
The consistent performance across modalities further validates the adaptability and resilience of \name in diverse and dynamic environments.
Moreover, we notice in Fig.~\ref{fig:dynamic_weight_noise1} and Fig.~\ref{fig:dynamic_weight_modality1} that although abrupt changes in physiological signals lead to a notable increase in input noise levels, the stability of estimated fluctuation noise adaptively regularizes the modality fusion weights, which helps the model avoid misinterpreting critical health biomarkers as irrelevant noise. 
In contrast, consistent input noise leads to a gradual decrease of the modality fusion weights to reduce reliance on unreliable modalities.
This balance between input and fluctuation noise enables our \name to remain sensitive to meaningful biomarker fluctuations while suppressing irrelevant disturbances.
}

% \begin{figure}[t]
% \vspace{-.5ex}
% % \setlength\abovecaptionskip{3pt}
% \centering
% \subfloat[Facial information.]{
%     \includegraphics[width=0.495\linewidth]{example-image-duck}
%     \label{fig:dynamic_weight_facial_acc}
% }
% \subfloat[modality-wise fusion weights.]{
%     \includegraphics[width=0.495\linewidth]{example-image-duck}
%     \label{fig:dynamic_weight_facial_weight}
% }  \\
% \subfloat[Physiological signal.]{
%     \includegraphics[width=0.495\linewidth]{example-image-duck}
%     \label{fig:dynamic_weight_phy_acc}
% }
% \subfloat[modality-wise fusion weights.]{
%     \includegraphics[width=0.495\linewidth]{example-image-duck}
%     \label{fig:dynamic_weight_phy_weight}
% }
% \caption{Health event predictions and modality-wise fusion weights.}
% \label{fig:dynamic_weight}
% \vspace{-2ex}
% \end{figure}

\begin{table}[t]
\centering
% \scalebox{0.9}{
  \begin{tabular}{c|cccc}
    \hline
    Method & Accuracy & Precision & Recall & F1 score \\
    \hline
    \texttt{a)} & 60.23 & 68.86 & 61.91 & 58.13 \\
    \texttt{b)} & 69.54 & 76.03 & 68.02 & 66.04 \\
    \texttt{c)} & 72.48 & 85.43 & 70.32 & 69.14 \\
    \texttt{d)} & 76.09 & 88.96 & 72.61 & 73.69 \\
    \texttt{e)} & \textbf{79.89} & \textbf{94.47} & \textbf{76.08} & \textbf{77.35} \\
    \hline
  \end{tabular}
% }
\caption{Ablation study of \name on uncertainty quantification and calibration.}
\vspace{-2ex}
\label{tab:ablation_uncertainty}
\end{table}

\vspace{-0.1cm}
\subsection{Ablation Study}
\begin{figure}[t]
\vspace{-.5ex}
\centering
\subfloat[Uncertainty quantification and \\calibration.]{\includegraphics[width=0.49\linewidth]{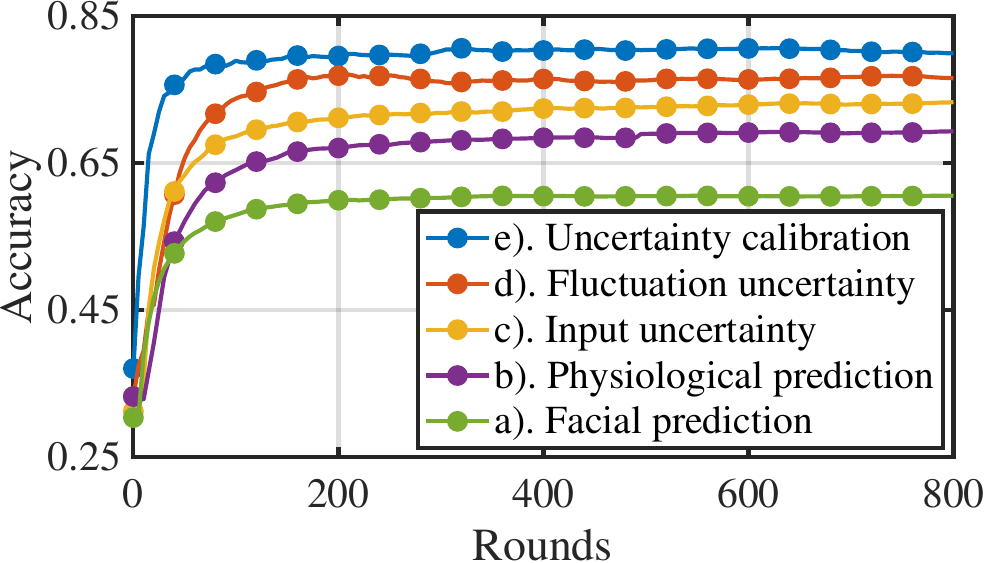}
    \label{fig:ablation_uncertainty_time}
}
\subfloat[Transformer-based multimodal \\fusion.]{\includegraphics[width=0.49\linewidth]{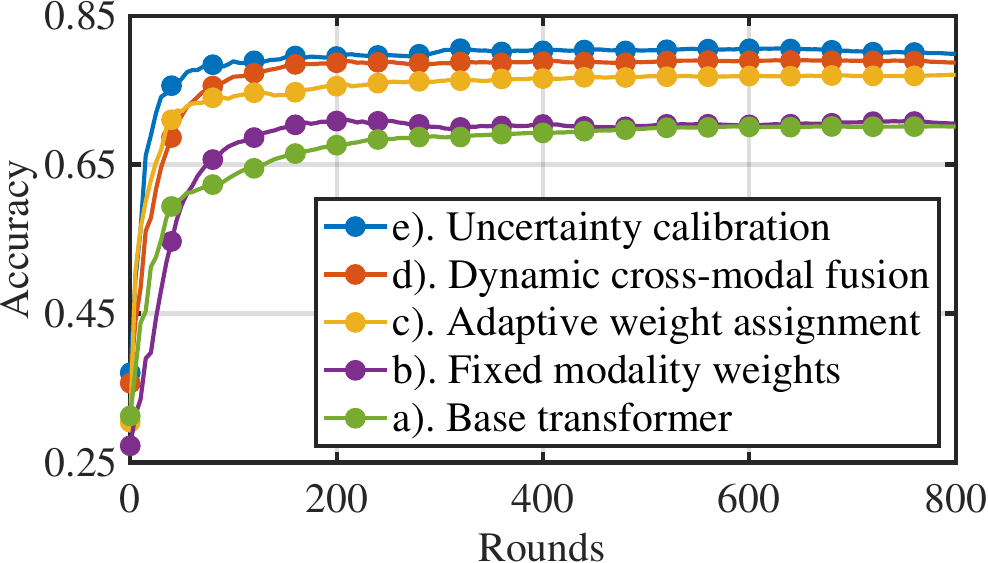}
    \label{fig:ablation_transformer_time}
}\\
\subfloat[Facial information missing.]{\includegraphics[width=0.49\linewidth]{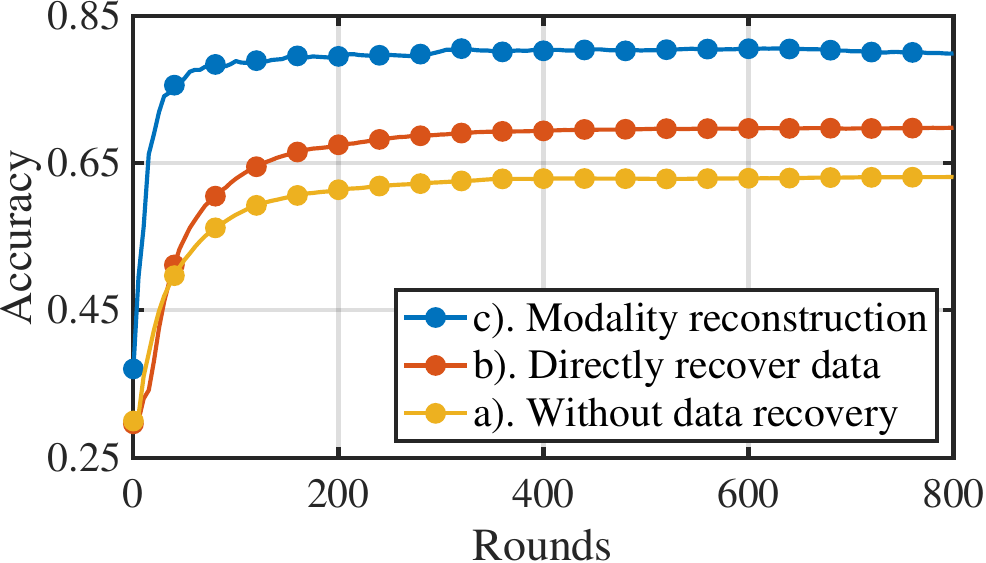}
    \label{fig:ablation_missing_facial}
}
\subfloat[Physiological signal missing.]{\includegraphics[width=0.49\linewidth]{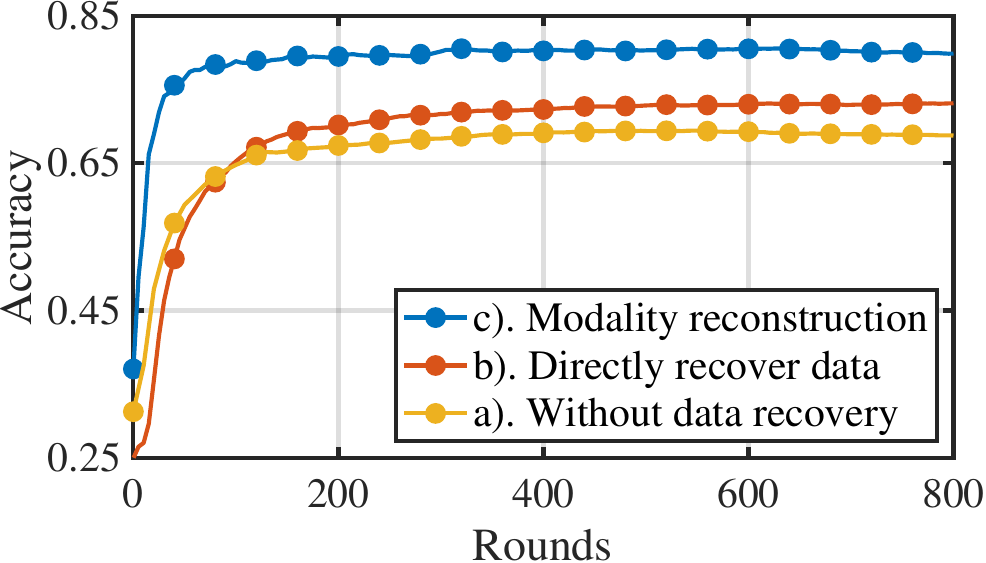}
    \label{fig:ablation_missing_physio}
}
\caption{Ablation study on the Stressors dataset.}
% \label{fig:robustness_modality}
\vspace{-2ex}
\end{figure}

{\textit{1) \textbf{Uncertainty Quantification and Calibration.}}}
Fig.~\ref{fig:ablation_uncertainty_time} and Table~\ref{tab:ablation_uncertainty} illustrate the impact of uncertainty quantification and calibration on the Stressors dataset with 50\% noisy inputs and 50\% facial information missing.
\waitrev{The accuracy of individual modalities when used independently presents their standalone contributions, revealing that some modalities carry more discriminative information than others for specific health status detections, underscoring the importance of properly handling modality uncertainty to mitigate the negative impact of low-quality noisy modalities}.
% However, performance improvement is limited by the failure to account for high-variance data fluctuations in uncertainty modeling, which misclassifies critical health indicators as noise. On the contrary, 
By quantifying input uncertainty arising from dynamic environments, the performance improves in both detection accuracy and precision.
However, the improvement is still limited by the failure to account for biomarker fluctuations in uncertainty modeling, which often result in the misclassification of critical health biomarkers as noise, \rev{restricting the F1 score lower than 70\%}.
By accounting for both input uncertainty and fluctuation uncertainty into adaptive weight assignment for multimodal fusion, the proposed approach achieves timely detection of health changes even under severe data degradation, reaching a remarkable \rev{76\% accuracy and 72\% recall}. 
\rev{Incorporating the calibration of modality contribution into model training brings further benefits \waitrev{as it allows model to focus on the most informative features}, guaranteeing the effectiveness of our \name for health monitoring with the proposed dynamic multimodal fusion module in dynamic driving environments.}

\begin{table}[t]
\centering
% \scalebox{0.9}{
  \begin{tabular}{c|cccc}
    \hline
    Method & Accuracy & Precision & Recall & F1 score \\
    \hline
    \texttt{a)} & 70.17 & 78.45 & 68.72 & 65.94 \\
    \texttt{b)} & 70.52 & 78.89 & 68.36 & 65.41 \\
    \texttt{c)} & 77.04 & 89.06 & 72.55 & 73.75 \\
    \texttt{d)} & 78.68 & 93.90 & 73.91 & 75.62 \\
    \texttt{e)} & \textbf{79.89} & \textbf{94.47} & \textbf{76.08} & \textbf{77.35} \\
    \hline
  \end{tabular}
% }
\caption{Ablation study of \name on transformer-based multimodal fusion.}
  \label{tab:ablation_transformer}
  \vspace{-1.5ex}
\end{table}

{\textit{2) \textbf{Transformer-based Multimodal Fusion.}}}
Fig.~\ref{fig:ablation_transformer_time} and Table~\ref{tab:ablation_transformer} present the impact of multimodal fusion in the transformer framework on the Stressors dataset with 50\% noisy inputs and 50\% facial information missing.
The results reveal the limitations of using fixed modality weights, which struggle to distinguish between useful signals and irrelevant information in multimodal data of varying quality. In contrast, dynamically assigning modality weight to prioritize reliable modalities demonstrate significant improvements, enhancing the \rev{accuracy by 6.5\% and precision by 10\%}. 
Moreover, compared to the base transformer model with 50\% noisy inputs, which treats all modalities equally in the self-attention mechanism, our proposed transformer-based multimodal fusion module improves precision by nearly \needrev{5\%}. This underscores the benefits of integrating dynamic cross-modal fusion into the transformer framework, which adjusts the attention of each modality to better capture critical cross-modal correlations, thereby enhancing robustness and accuracy in outdoor health monitoring in dynamic environments.

\begin{table}[t]
\centering
\scalebox{0.98}{
  \begin{tabular}{c|cccc}
    \hline
    Method & Accuracy & Precision & Recall & F1 score \\
    \hline
    \texttt{Facial a)} & 58.94 & 67.20 & 60.16 & 56.74 \\
    \texttt{Physio a)} & 68.79 & 69.76 & 63.08 & 60.21 \\
    \texttt{Facial b)} & 69.78 & 75.14 & 68.18 & 65.72 \\
    \texttt{Physio b)} & 73.13 & 85.57 & 69.87 & 70.29 \\
    \texttt{\name} & \textbf{79.89} & \textbf{94.47} & \textbf{76.08} & \textbf{77.35} \\
    \hline
  \end{tabular}
}
\caption{Ablation study of \name on missing modality reconstruction.}
  \label{tab:ablation_missing}
  \vspace{-2ex}
\end{table}

{\textit{3) \textbf{Missing Modality Reconstruction.}}}
Fig.~\ref{fig:ablation_missing_facial}, Fig.~\ref{fig:ablation_missing_physio}, and Table~\ref{tab:ablation_missing} compare the impact of missing modality reconstruction on the Stressors dataset with 50\% noisy inputs and 50\% missing samples in different modalities.
% b). directly recover data under high data quality of other modalities
We evaluate the performance of modality reconstruction by comparing model training under three conditions: a). modality reconstruction with distribution normalization under varying data degradation of other modalities, b). directly recover data with varying quality of other modalities, and c). modality missing without data recovery. 
Our proposed modality reconstruction module outperforms the model with direct data recovery, achieving nearly \needrev{7\%} improvement in accuracy. This is owing to the advantages of normalizing modality distribution into the common space, \rev{which allows the model to learn consistent relationships, thus facilitating the extraction of stable cross-modal correlations despite distribution fluctuations.}
It is also worth noting that the performance of different missing modalities are comparable, showing the robustness of \name to learn the consistent cross-modal correlations with dynamic changing environment and perform accurate modality reconstruction.

% % different modalities missing: before and after predict @ average precision and recall (importance of different modalities and the recovery performance)(2 Figure)
% We further show \newrev{the robustness of \name in cross-domain setting by investigating how the trained model performs in different modalities missing}. The model is trained with 90\% of a single modality (camera, wearable sensors or vehicle data) missing under different labeling rate and the testing accuracy is reported in Fig.~\ref{fig:robustness_modality_1}.
% As Fig.~\ref{fig:robustness_modality_statistics_1} illustrates, \name is still able to detect most of the abnormal status under various modality missing, with the accuracy achieving \needrev{xx\%, xx\% and xx\% for camera, wearable sensors and vehicle data missing respectively}. We also observe that the difference of the performance under different settings is \needrev{negligible}, confirming that \name has the ability to learn the correlations between modalities and perform accurate modality reconstruction.
% \needrev{Moreover, it is clear in Fig.~\ref{fig:robustness_modality_facial_1}, \ref{fig:robustness_modality_vital_1} and \ref{fig:robustness_modality_vehicle_1} that the accuracy increases as the labeling rate rises, however, it is worth noting that the tendency grows gradually slower as \rev{the number of labelled date turns to be sufficient for training \name}, which validates the advantage \name has in the dataset with only small portion of labelled data.}

% \vspace{-0.1cm}
\section{Conclusion} \label{sec:conclusion}
In this paper, we have proposed an uncertainty-aware multimodal fusion framework, named \name, for outdoor health monitoring in dynamic and noisy environments. 
% to achieve dynmaic health monitoring under changing environments with varying data quality, aimed at enhancing accuracy and timeliness of health status detection. 
% We have first determined modality-specific fusion weights by quantifying uncertainty from both input reliability and fluctuation sensitivity to enhance feature extraction and uncertainty estimation. 
We have first quantified modality uncertainty caused by input and fluctuation noise utilizing current and temporal features.
We have then introduced a transformer-based multimodal fusion to determine modality-specific fusion weights based on modality uncertainty with calibrated unimodal contribution, enhancing the detection of critical cross-modal relationships in the presence of low-quality data.
\rev{Finally, we have designed a missing modality reconstruction network that maps fluctuating modality distributions into a common space, facilitating stable cross-modal alignment for accurate data recovery.}
Extensive experiments have demonstrated that our \name framework achieves superior performance compared to the state-of-the-art baselines. As a potential future direction, we are looking forward to
 extending our \name to improve the performance of various applications such as
 distributed learning systems~\cite{lin2024efficient,hu2024accelerating,zhang2025lcfed,lin2024adaptsfl,zhang2024fedac}.

% \section{Acknowledgments}
% \needrev{This work was supported in part by the Hong Kong SAR Government under the Global STEM Professorship and Research Talent Hub, in part by the Hong Kong Jockey Club under the Hong Kong JC STEM Lab of Smart City (Ref.: 2023-0108), and in part by the Hong Kong Innovation and Technology Commission under InnoHK Project CIMDA. The work of Yiqin Deng was supported in part by the National Natural Science Foundation of China under Grant No. 62301300. The work of Xianhao Chen was supported in part by HKU-SCF FinTech Academy R\&D Funding.}

\ifCLASSOPTIONcaptionsoff
  \newpage
\fi



%\bibliographystyle{IEEEtran}


% \end{thebibliography}

\bibliographystyle{IEEEtran}
\bibliography{reference}

\end{document}